\documentclass[letter]{aa}

\usepackage{graphicx}
\usepackage{amsmath}
\usepackage{amssymb}
\usepackage{comment}
\usepackage{hyperref}
\usepackage{subfigure}
\usepackage{ulem}

\usepackage[varg]{txfonts}
\usepackage{booktabs}
\usepackage{xcolor}

\newcommand{\kms}{km s$^{-1}$}
\newcommand{\myr}{$M_{\odot}$ yr$^{-1}$}
\newcommand{\msun}{$M_{\odot}$}
\newcommand{\mstar}{$M_{\star}$}
\newcommand{\mgas}{$M_{\rm gas}$}
\newcommand{\mdust}{$M_{\rm dust}$}

\newcommand{\kkmspclsun}{K~km~s$^{-1}$~pc$^2$~$L_\odot^{-1}$}
\newcommand{\jykms}{Jy~km~s$^{-1}$}
\newcommand{\ci}{[C\,{\footnotesize I}]}

\newcommand{\co}{CO}
\newcommand{\cione}{[C\,{\footnotesize I}]$(^3P_1\,-\,^{3}P_0)$}
\newcommand{\citwo}{[C\,{\footnotesize I}]$(^3P_2\,-\, ^{3}P_1)$}

\newcommand{\cofive}{CO\,$(5-4)$}

\newcommand{\coseven}{CO\,$(7-6)$}

\newcommand{\cothree}{CO\,$(3-2)$}

\newcommand{\coone}{CO\,$(1-0)$}

\newcommand{\lprimecothree}{$L'_{\rm CO(3-2)}$}

\newcommand{\lprimecofive}{$L'_{\rm CO(5-4)}$}
\newcommand{\lprimecoseven}{$L'_{\rm CO(7-6)}$}

\newcommand{\lcoseven}{$L_{\rm CO(7-6)}$}

\newcommand{\lir}{$L_{\rm IR}$}

\newcommand{\lsun}{$L_{\odot}$}

\newcommand{\jwst}{\textit{JWST}}

\newcommand\tcb[1]{\textcolor{black}{#1}}

\newcommand{\ciones}{[CI]\,($1-0$)}
\newcommand{\citwos}{[CI]\,($2-1$)}

\begin{document} 
   
   \title{Molecular gas content and high excitation of a massive main-sequence galaxy at $ z = 3 $}

   \author{Han Lei
          \inst{1,2}
          \and
          Francesco Valentino\inst{1,2,3}
          \and
          Georgios E. Magdis\inst{1,2,4}
          \and
          Vasily Kokorev\inst{5}
          \and
          Daizhong Liu\inst{6}
          \and
          Dimitra Rigopoulou\inst{7,8}
          \and
          Shuowen Jin\inst{1,4}
          \and
          Emanuele Daddi\inst{9}
          }

   \institute{Cosmic Dawn Center (DAWN), Denmark
             \and
             Niels Bohr Institute, University of Copenhagen, Jagtvej 128, 2200 Copenhagen N, Denmark
             \and
             European Southern Observatory, Karl-Schwarzschild-Str. 2, D-85748 Garching bei M\"{u}nchen, Germany
             \and 
             DTU-Space, Technical University of Denmark, Elektrovej 327, DK-2800 Kgs. Lyngby, Denmark
             \and
             Kapteyn Astronomical Institute, University of Groningen, PO Box 800, 9700 AV Groningen, Netherlands
             \and
             Max-Planck-Institut f\"{u}r extraterrestrische Physik (MPE), Giessenbachstrasse 1, D-85748 Garching, Germany
             \and
             Astrophysics, Department of Physics, University of Oxford, Keble Road, Oxford OX1 3RH, UK 
             \and
             School of Sciences, European University Cyprus, Diogenes Street, Engomi, 1516 Nicosia, Cyprus
             \and
             CEA, IRFU, DAp, AIM, Universit\'{e} Paris-Saclay, Universit\'{e} de Paris, CNRS, F-91191 Gif-sur-Yvette, France
             }
  
  \abstract{  
 We present new \co\ ($J=5-4$ and $7-6$) and \ci\ ($^3P_2\,-\, ^3P_1$ and $^3P_1\,-\, ^3P_0$) emission line observations of the star-forming galaxy D49 at the massive end of the main sequence at $z=3$. We incorporate previous \co\ ($J=3-2$) and optical-to-millimetre continuum observations to fit its spectral energy distribution (SED). Our results hint at high-$J$ CO luminosities exceeding the expected location on the empirical correlations with the infrared luminosity. [CI] emission fully consistent with the literature trends is found. We do not retrieve any signatures of a bright active galactic nucleus that could boost the $J=5-4,\,7-6$ lines in either the infrared or X-ray bands, but warm photon-dominated regions, shocks, or turbulence could in principle do so. We suggest that mechanical heating could be a favourable mechanism able to enhance the gas emission at fixed infrared luminosity in D49 and other main-sequence star-forming galaxies at high redshift, but further investigation is necessary to confirm this explanation. We derive molecular gas masses from dust, \co, and \ci\ that all agree within the uncertainties. Given its high star formation rate (SFR) $\sim500$ \myr\ and stellar mass $>10^{11.5}$ \msun, the short depletion  timescale of $<0.3$ Gyr might indicate that D49 is experiencing its last growth spurt and will soon transit to quiescence.}


   \maketitle

\section{Introduction}

Star formation plays an essential role in the evolution of galaxies. It is the driver of their stellar mass (\mstar) growth; it leads to the chemical enrichment of the interstellar and circumgalactic media (ISM, CGM) via supernova explosions and stellar wind ejections; and its cessation determines the death of galaxies and their final transition to quiescence. It is now well established that the ensemble of galaxies in the Universe forms stars at a pace that, in terms of cosmic star formation rate (SFR) density, peaks around $z\sim2-3$, the period of maximum activity dubbed `cosmic noon' \citep{MadauDickinson2014}. Interestingly, most star-forming galaxies seem to follow tight correlations between \mstar and SFR, which is known as the main sequence of star formation \citep{Daddi2007, Elbaz2007, Noeske2007}, and between the surface densities of gas mass ($\Sigma_{\rm gas}$) and SFR ($\Sigma_{\rm SFR}$), the Schmidt-Kennicutt relation \citep{Kennicutt1998}. \tcb{Only a small percentage of galaxies undergoing intense star formation activity (dubbed “starbursts”) seem to fall out of the correlations} \citep{Daddi2010, Genzel2010mn, Rodighiero2011, Rodighiero2014,Tacconi2020}. Accurate estimates of the amount and properties of the molecular hydrogen reservoirs ($\mathrm{H}_2$), the primary fuel for the formation of new stars, therefore become the key to understanding how galaxies evolve across cosmic time.\\ 

The lack of dipole moment and the warm temperatures ($\sim500\,\mathrm{K}$) necessary to excite the first rotational transition make $\mathrm{H}_2$ difficult to be directly detected in galaxies. Thus, alternative proxies are used to trace the bulk of their cold molecular gas and measure its mass ($M_{\rm gas}$). Carbon monoxide (CO), the second most abundant molecule after $\mathrm{H}_2$, is a traditional choice. However, its first rotational transition CO($J=1-0$) ($\nu_{\rm rest} = 115.27\,{\rm GHz}$; hereafter \coone ) is difficult to observe for galaxies at the peak of star formation activity at cosmic noon ($z\sim2-3$). Moreover, the conversion from the CO luminosity to $M_{\rm gas}$ depends on the physical conditions of local ISM \citep{Carilli2013}. Optically thin dust emission has emerged as a reliable and observationally inexpensive alternative. \tcb{\mgas\ can be estimated by multiplying dust masses (\mdust) from the modelling of the far-infrared spectral energy distribution (SED) by a metallicity-dependent gas-to-dust ratio \citep{Magdis2011, Magdis2012, Berta2016}. In alternative, \mgas\ can be derived directly from single sub-millimetre band observations calibrated against CO \citep{Scoville2014,Scoville2016,Liu2019_supplementary, Liu2019_letter, Kaasinen2019}.} In the past, the \cione\ and \citwo\ transitions of neutral atomic carbon ($\nu_{\rm rest} = 492.16\,{\rm GHz}$ and $809.34\,{\rm GHz}$, respectively; hereafter \ciones\, and \citwos) were brought forward to serve as potentially robust tracers of \mgas, and were  even better than the traditional CO under certain circumstances \citep[e.g.][]{Papadopoulos2004, Papadodoulos2004b, Madden2020, Dunne2022MNRAS.tmp.1998D}. Being optically thin even in the most extreme star-forming environment \citep{Harrington2021}, \ci\ line fluxes can be converted  
to \mgas\ with an assumption of the gas excitation and an abundance \citep{Papadodoulos2004b}. Furthermore, the combination of several CO and [CI] transitions can constrain the properties and plausible gas heating mechanisms in galaxies  \citep[e.g.][]{Bothwell2017,  Harrington2021, Liu2021, Papadopoulos2022}. However, due to their intrinsic brightness, observations and modelling of typical galaxies on the main sequence of star formation at cosmic noon have been explored to a much lesser extent compared to their starbursting counterparts \citep{Valentino2020, Brisbin2019, Bourne2019, Boogaard2020, Henriquez2022}.\\

Here we present a comprehensive study of the massive star-forming galaxy D49 located at the high-mass end of the main sequence at $z=2.84$ \citep{Magdis2017}. This object is spectroscopically confirmed via Lyman-$\alpha$ \citep{Steidel2003, Shapley2003} and \cothree\ \citep{Magdis2017}. Moreover, its SED is finely sampled from the optical to the millimetre regime. This makes D49 a perfect test-bed for the calibration of different gas tracers and the exploration of the molecular gas properties on the main sequence, possibly during the last epoch of assembly before quenching.

In this letter we present new measurements of the \cofive, \coseven, \citwos, and \ciones\ emission lines and their underlying dust continuum using the NOrthern Extended Millimetre Array (NOEMA).
When estimating gas masses, \tcb{we assume  $M_{\rm gas} \approx M_{\rm H_2}$ to be consistent with the  \cite{Magdis2017} values.} We adopt throughout ${\rm \Omega_m} = 0.3$, ${\rm \Omega_\Lambda} = 0.7$, $H_0 = 71\,{\rm km\,s^{-1}\,Mpc^{-1}}$, and a \cite{Chabrier2003} IMF.

\section{Observations}
D49 is a massive and infrared-bright galaxy, originally identified from the optically selected (\textit{U, G, R}) sample of Lyman-break galaxies (LBGs) in the Extended Groth Strip field (EGS) at $z\sim3$ in \cite{Steidel2003}. We observed D49 with NOEMA (W18DT, PI: I. Cortzen). The antennas were pointed at (RA, Dec) = (14h:17m:29s.234, 52$^\circ$:34$^\prime$:34$^{\prime\prime}$.450). The galaxy redshift $z = 2.846$ is known from previous detections of the CO(3-2) line \citep{Magdis2017} and from optical spectroscopy described in \cite{Steidel2003}. Therefore, the expected frequencies of \coseven, + \citwos, and \cofive, + \ciones\ are 209.74+210.43 GHz and 149.84 GHz+127.97 GHz, respectively, falling in Bands 3 and 2. We observed D49 in each band for 2.5 hours in D configuration on March 28 and April 13, 2019. The data were calibrated by the standard pipeline contained in the {\ttfamily GILDAS} package.\footnote{\url{https://www.iram.fr/IRAMFR/GILDAS}} In the images cleaned with default recipes with \texttt{MAPPING} \citep{Hogbom1974}, the final beam sizes are $1.93^{\prime\prime} \times 1.48^{\prime\prime}$ (${\rm PA} = 62^\circ$) and $3.19^{\prime\prime} \times 2.50^{\prime\prime}$ (${\rm PA} = 63^\circ$) at Band3/1.3 mm and Band 2/2 mm, respectively. The spectra were extracted with circular Gaussian profiles centred at the location of the \cothree\, emission and with a fixed full width at half maximum (FWHM) of $0.6^{\prime\prime}$ ($= 4.8\,{\rm kpc}$ at $z=2.84$) to be consistent with the results in \cite{Magdis2017}. The extraction was performed in the \textit{u-v} space with the task \texttt{UV\textunderscore FIT}. 
The continuum emission was extracted in a similar way after averaging the line-free spectral windows. Given the spatial resolution, we did not resolve the source. This is further confirmed by the consistent continuum flux densities extracted by fitting the emission with a circular Gaussian, a point source profile, and by reading the brightest pixel in the cleaned image.

\section{Results}

\subsection{CO and [CI] emission lines}

\begin{figure}
   \centering
   \includegraphics[width=\columnwidth,trim={0.1cm 0.3cm 0.5cm 0.5cm},clip]{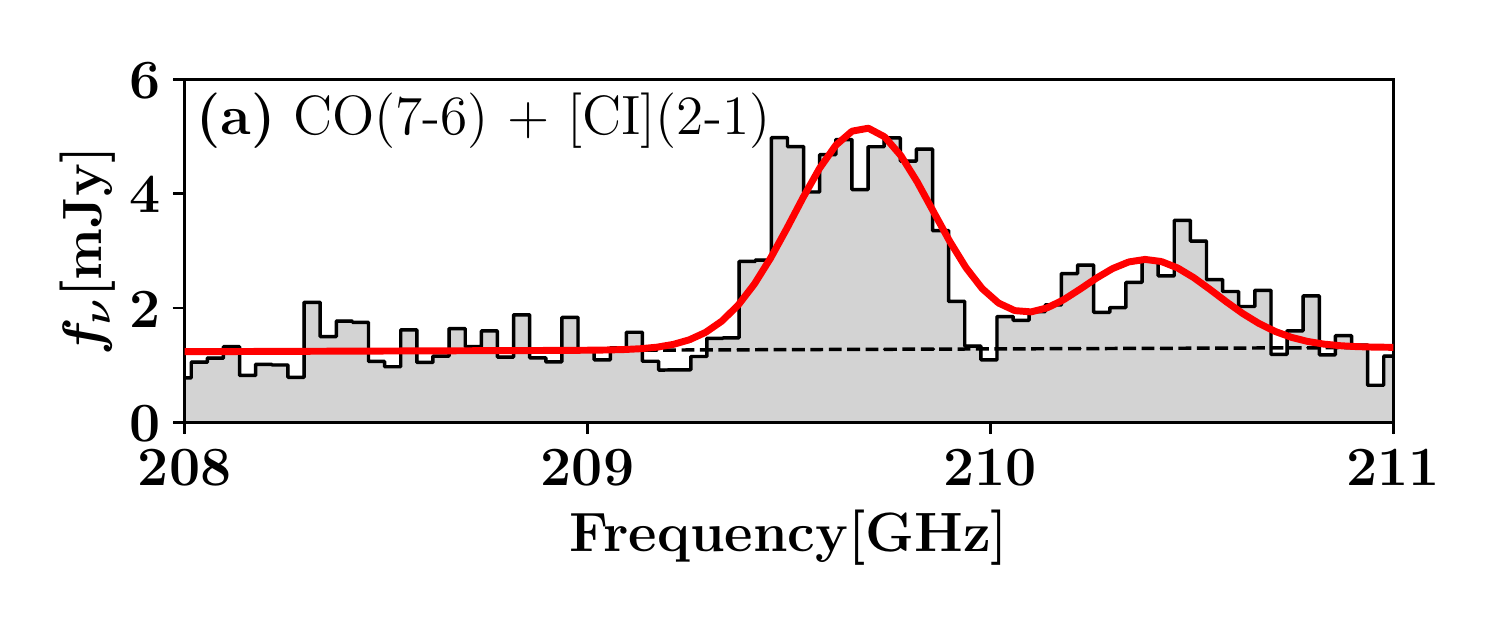}
   \includegraphics[width=\columnwidth,trim={0.1cm 0.3cm 0.5cm 0.5cm},clip]{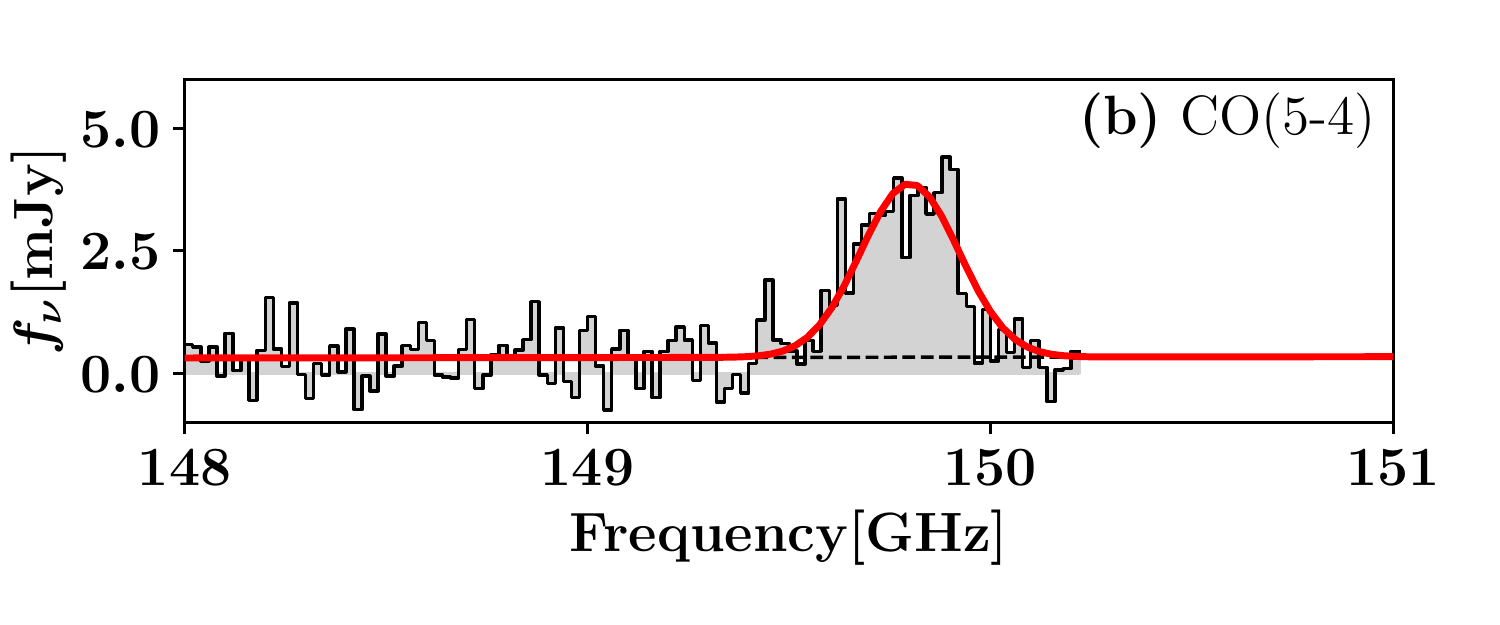}
   \includegraphics[width=\columnwidth,trim={0.1cm 0.3cm 0.5cm 0.5cm},clip]{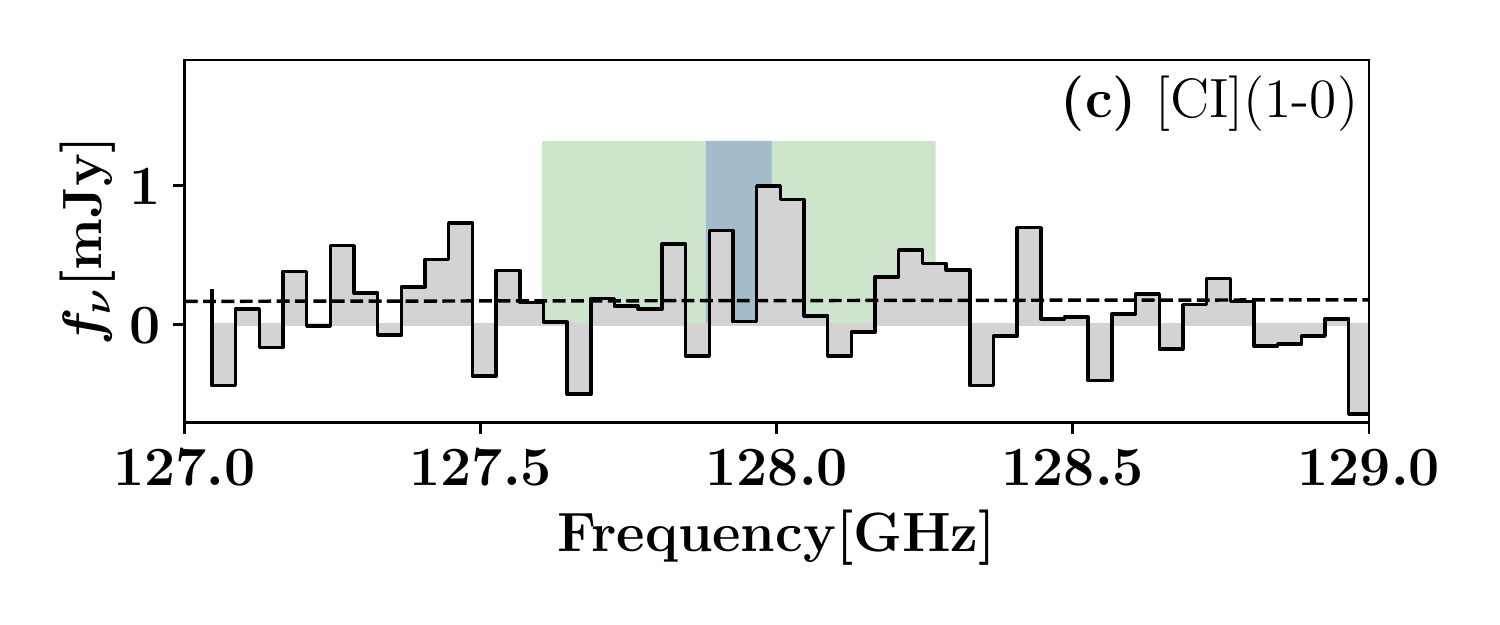}
   \caption{Spectra of (a) \coseven\, + \citwos\,, (b) \cofive\,, and (c) \ciones\, emission of D49. The binned step width is $\sim 90 \, {\rm km\,s^{-1}}$ in panels (a) and (c) and $\sim 40 \, {\rm km\,s^{-1}}$ in panel (b). The red curves indicate the fitted Gaussian profiles combined with a power-law dust continuum, the latter shown by a dashed line. The blue and green shaded areas in panel (c) indicate the $1\sigma_v$ and $\pm3\sigma_v$ velocity width used to set the upper limit on \ciones.}
   \label{fig.specs}
\end{figure}

In the extracted spectra shown in Fig. \ref{fig.specs}, we fit Gaussian profiles for lines and a power-law dust continuum to the observed data. We fixed the redshift and line widths of the three clearly detected lines (\cofive\, at $14\sigma$, \coseven\, at $16\sigma$, and \citwos\, at $8\sigma$), assuming that the molecules and atoms producing them belong to the same regions of the unresolved galaxy. The best-fit results are listed in Table \ref{tab.prop}. The best-fit redshift and line FWHM are $(2.8469 \pm 0.0001)$ and $(590\pm25)$ \kms, which are similar to those in \text{\cite{Magdis2017}}. We then derived $L'$ line luminosities as in \cite{Solomon2005}:
\begin{eqnarray}
L'_{\rm line}\,{\rm [K\,km\,s^{-1}\,pc^2]} = 3.25 \times 10^7 S_{\rm line}\Delta v \, \nu_{\rm obs}^{-2} D_L^2(1+z)^{-3}. 
\end{eqnarray}
Here $S_{\rm line}\Delta v$ is the integrated line flux in ${\rm Jy\,km\,s^{-1}}$, $D_L$ is the luminosity distance in ${\rm Mpc}$, and $\nu_{\rm obs}$ is the observed frequency of the emission line in GHz. We also fit the spectra without fixing line width and redshift and the results are fully consistent within the uncertainties.

We did not detect any significant \ciones\, emission line (Fig. \ref{fig.specs}). We thus set an upper limit following Eq. (7) in \cite{Bothwell2013},
\begin{eqnarray}
   S_{\rm [CI](1-0)}\Delta v < 3\times {\rm RMS}_{\rm channel} \times \sqrt{6\sigma_v {\rm d}v},
   \label{eq.ul}
\end{eqnarray}
where ${\rm RMS}_{\rm channel}$ is the root mean square per channel estimated locally around the expected location of the line, $\sigma_v$ is the line width that we impose to be identical to that of all the other lines ($\mathrm{FWHM} \approx 2.355\times\sigma_v$), and ${\rm d}v$ is the channel velocity width. We computed a conservative upper limit within    $\pm 3 \sigma_v$  (Fig. \ref{fig.specs}), corresponding to the $\sqrt{6}$ factor in Eq. (\ref{eq.ul}). 

\begin{table}
      \caption{Best-fit results of CO and [CI] lines of D49, obtained by fixing the redshift and FWHM for \cofive, \coseven, and \citwos. The upper limit on \ciones\, is computed within   $ \pm 3\sigma_v$, with the best-fit redshift $z_{\rm CO(5-4)} = z_{\rm CO(7-6)} = z_{\rm [CI](2-1)} = 2.847$. \cothree\ is from \cite{Magdis2017}.
      }
        \label{tab.prop}
        $$
        \centering
        \begin{array}{lccc}
            \hline
            \hline
            \noalign{\smallskip}
             & \multicolumn{1}{c}{S\Delta v} & \multicolumn{1}{c}{\rm FWHM} & \multicolumn{1}{c}{\log L'}\\
            & \multicolumn{1}{c}{\rm [Jy\,km\,s^{-1}]} & \multicolumn{1}{c}{\rm [km\,s^{-1}]} & \multicolumn{1}{c}{\rm [K\,km\,s^{-1}\,pc^2]}\\
            \noalign{\smallskip}
            \hline
            \noalign{\smallskip}
            {\rm CO(\operatorname{3-2})} & 1.10 \pm 0.18 & 501 \pm 35  &  10.62 \pm 0.07  \\
            {\rm CO(\operatorname{5-4})} & 2.23 \pm 0.16 & 590 \pm 25  &  10.50 \pm 0.03  \\
            {\rm CO(\operatorname{7-6})} & 2.43 \pm 0.15 & 590 \pm 25  &  10.24 \pm 0.03  \\
            {\rm [CI](\operatorname{2-1})} & 0.98 \pm 0.12 & 590 \pm 25  &  9.84 \pm 0.05\\
            {\rm [CI](\operatorname{1-0})} & <0.44 & 590 & < 9.93\\
            \noalign{\smallskip}
            \hline
         \end{array}
     $$
\end{table}

\subsection{Spectral energy distribution modelling}
\label{sec:modeling}

\begin{figure}
   \centering
   \includegraphics[width=\columnwidth,trim={0.5cm 0.5cm 0.5cm 0.5cm},clip]{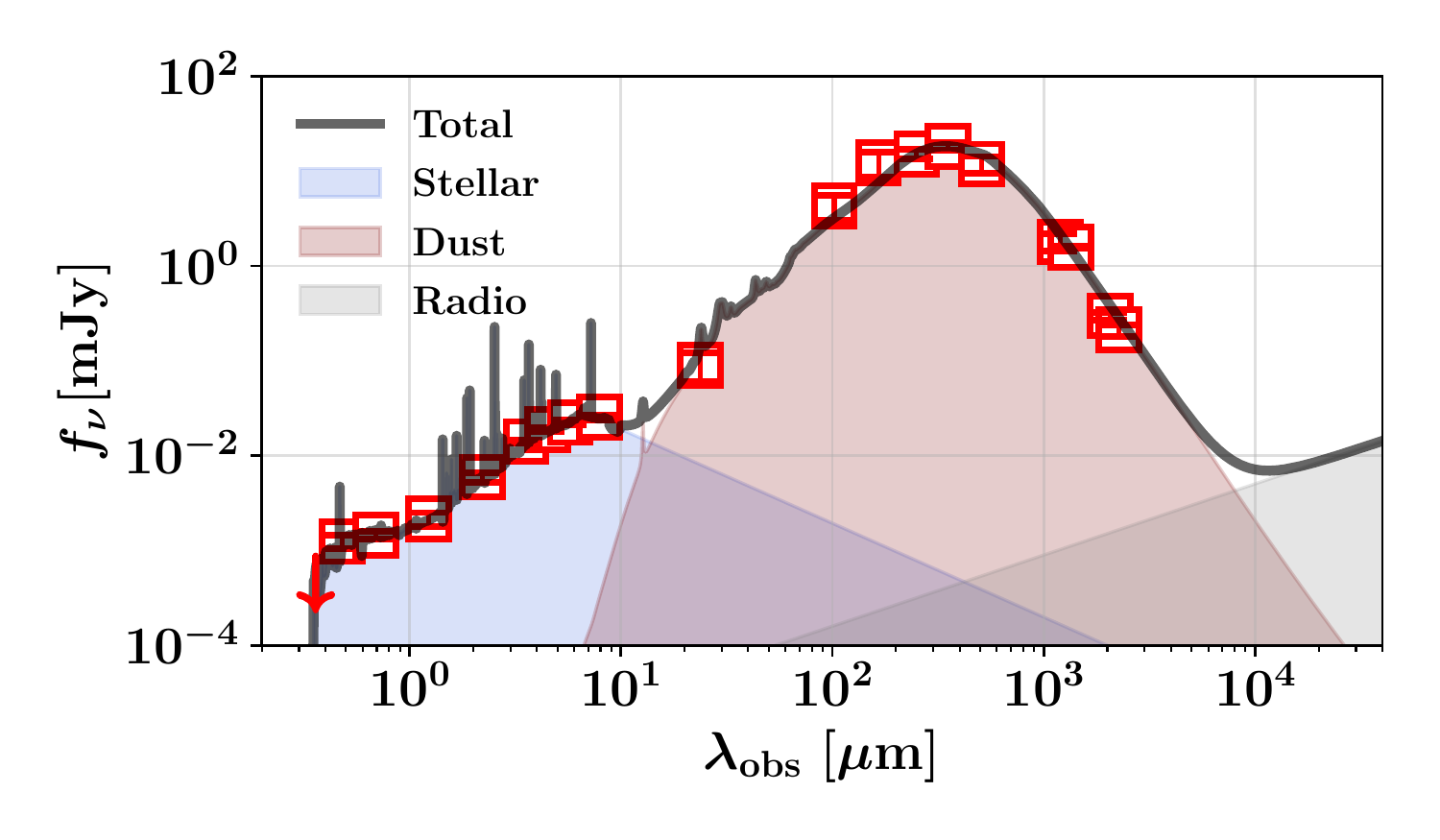}
   \caption{Spectral energy distribution of D49. The red open squares and downward arrow indicate the fitted photometry and upper limits. The black line shows the {\ttfamily Stardust} best-fit SED, while the contribution from stars and dust are respectively shown as a blue and red filled region. The grey region presents the predicted contribution from radio continuum, based on the radio-FIR model described in \cite{Delvecchio2021}.}
   \label{fig.full.spec}
\end{figure}

\begin{table}
    \caption{{\ttfamily Stardust} fitting result.}
    \label{tab.stardust}
    \centering
    \begin{tabular}{cc}
    \hline\hline            
    \noalign{\smallskip}
    $\mathrm{log}L_{\rm IR}[L_\odot]$ & $12.75 \pm 0.02$  \\
    $\mathrm{SFR} [M_\odot\,\mathrm{yr}^{-1}] $ & $566 \pm 24$  \\
    $\mathrm{log} M_{\rm dust} [M_\odot]$ & $9.27 \pm 0.03$  \\
    $\mathrm{log} M_{\rm gas} [M_\odot]$ & $11.19 \pm 0.03$  \\
    $\mathrm{log} M_\star [M_\odot]$ & $11.61 \pm 0.16$  \\
    Gas-to-dust ratio & $ 83 $  \\
    \hline
    \end{tabular}
\end{table}

 We modelled the full SED of D49 with the novel algorithm \href{https://github.com/VasilyKokorev/stardust}{\ttfamily Stardust} \citep{Kokorev2021}. {\ttfamily Stardust} can simultaneously, but independently fit linear combinations of stellar active galactic nuclei (AGN), and dust templates without explicitly imposing an energy balance between the absorbed \tcb{UV--optical} radiation and the IR emission. The latter is often assumed in SED modelling (e.g. {\ttfamily CIGALE}, \citealt{Burgarella2005, Noll2009, Boquien2019}; {\ttfamily MAGPHYS} \citealt{deCunha2008, Battisti2019}). This approach assumes that the stellar and dust emissions are spatially coincident, meaning that the UV absorption and the subsequent re-emission at IR wavelengths occur in the same environment \citep{deCunha2008}. However, resolved observations of distant galaxies at $z \sim 2$ have revealed offsets between the dust and stellar spatial distributions, challenging the energy balance assumption \citep{Chen2017ApJ...846..108C, Gabriela2018ApJ...863...56C, Cochrane2021MNRAS.503.2622C}. We thus relinquish this assumption in our models. For each component, we used the standard libraries released with {\ttfamily Stardust}: stellar population synthesis models in common with \texttt{eazy} from \cite{Brammer2008}, AGN templates in \cite{Mullaney2011}, and dust emission from \cite{DL07} and \cite{Draine2014}. The modelled photometric data, including the new dust continuum emission measurements presented in this work, are listed in Table \ref{tab.data} and the best-fit SED is shown in Fig. \ref{fig.full.spec}. We list the best-fit properties returned by {\ttfamily Stardust} in Table \ref{tab.stardust}. Even including the AGN templates, the best fit does not favour a meaningful solution with a strong AGN. Its possible contribution to the total IR budget is $f_{\rm AGN} = L_{\rm IR, AGN}/L_{\rm IR, total} = 0.01 \pm 0.33$, with ${\rm S/N} \ll 3$ on $L_{\rm IR, AGN}$). In other words, the emission can be fully accounted for with stellar and dust templates. We find that the infrared luminosity \lir($8-1000~{\rm \mu m}$) and dust mass \mdust\ are consistent with those in \cite{Magdis2017}, while {\ttfamily Stardust} returns 0.3 dex larger \mstar. We convert \lir\ into SFR as $L_{\rm IR}\, [L_\odot] = 10^{10} \times {\rm SFR}\, [M_\odot\,{\rm yr^{-1}}]$ (\citealt{Kennicutt1998rev} for a \citealt{Chabrier2003} IMF).
Assuming that D49 follows the fundamental metallicity relation \citep[FMR,][]{Mannucci2010}, we derive a gas-to-dust ratio of $\delta_{\rm GDR}(12+\mathrm{log(O/H)}=8.71\sim Z_\odot) \approx 83$ following \cite{Magdis2017}.

\section{Discussion}

\subsection{CO and [CI] emission compared with that of other galaxies}

\begin{figure*}
\centering
    \includegraphics[width=0.4\textwidth,trim=0.45cm 10 0.88cm 10,clip]{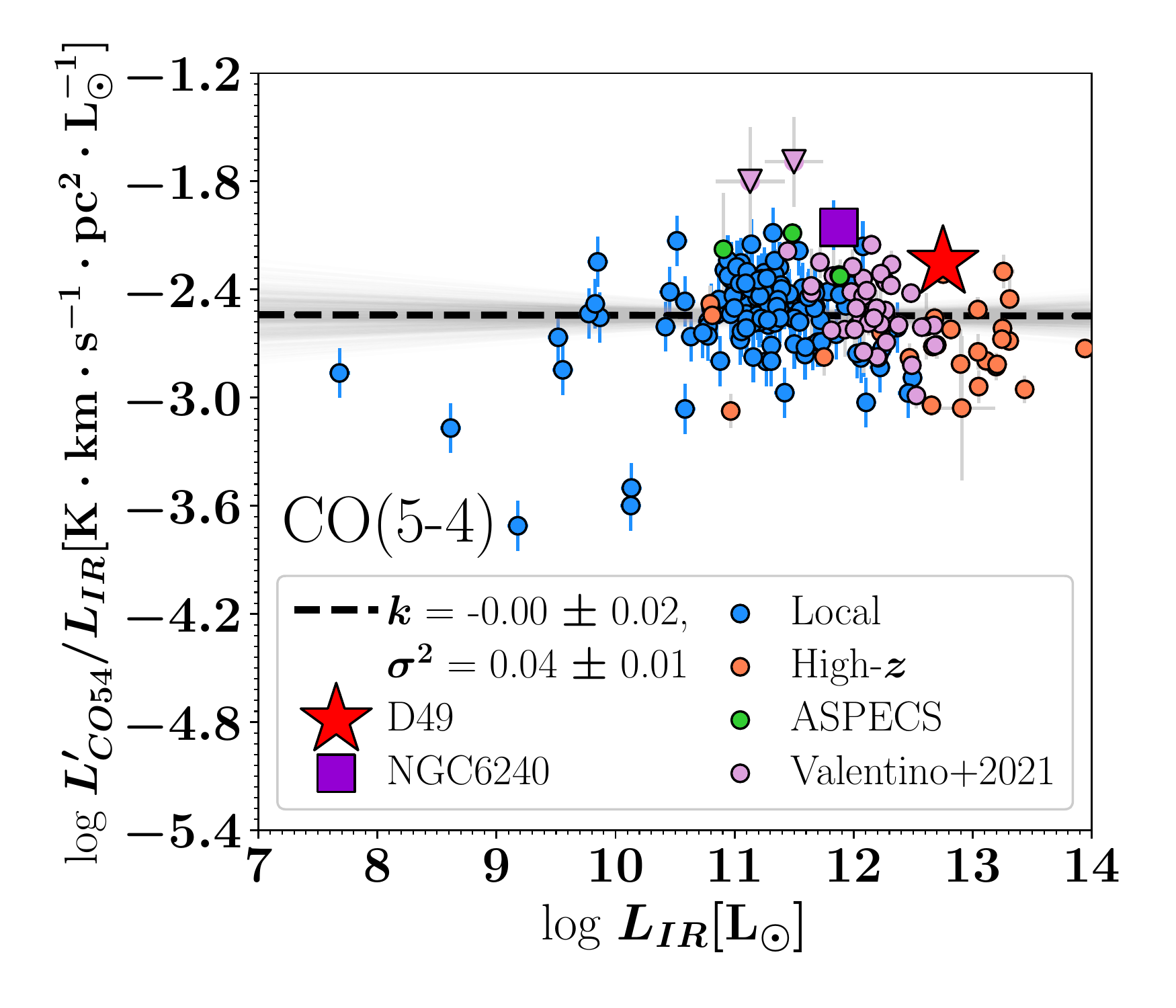}
    \quad
    \includegraphics[width=0.4\textwidth,trim=0.45cm 10 0.88cm 10,clip]{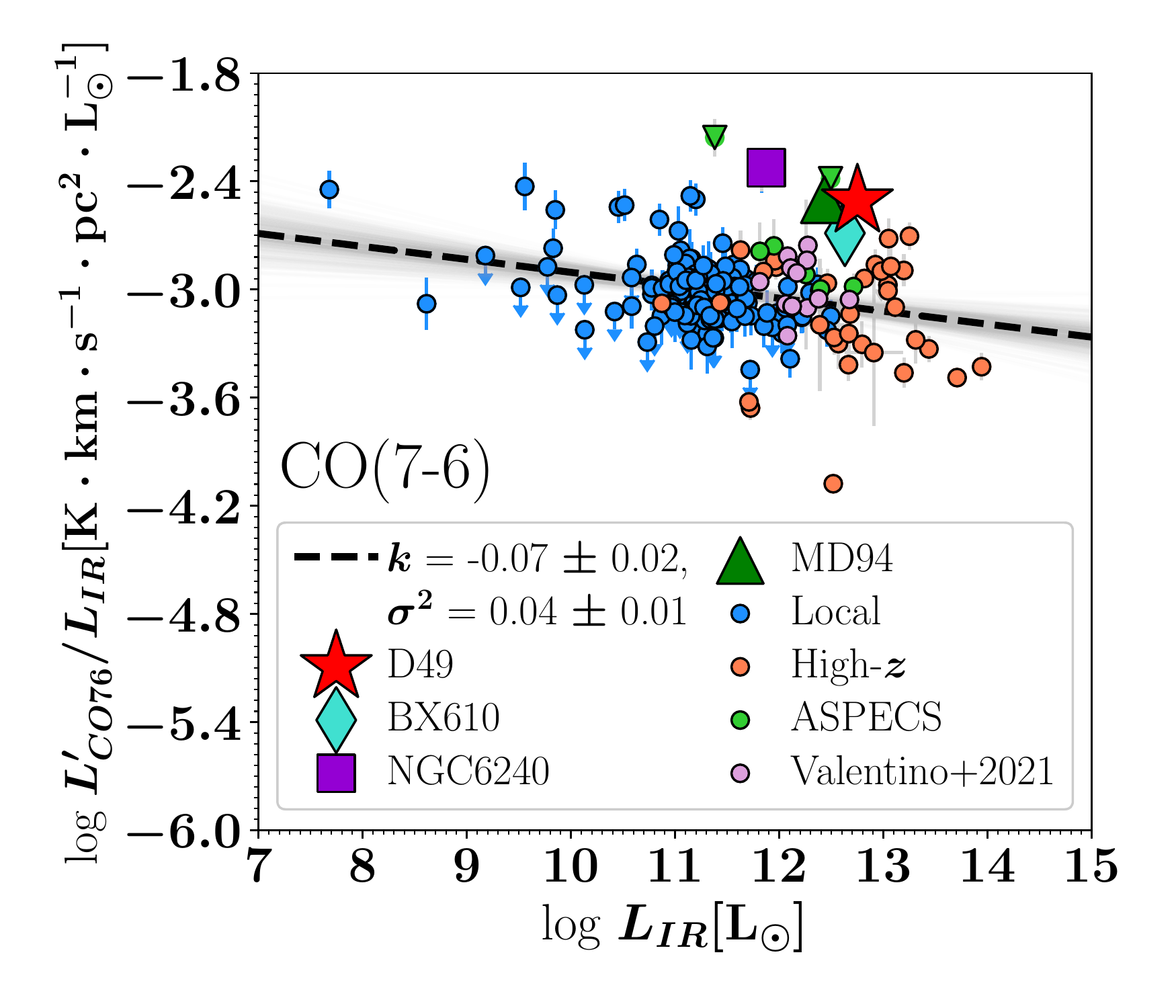}
    \includegraphics[width=0.4\textwidth,trim=0.45cm 10 0.88cm 10,clip]{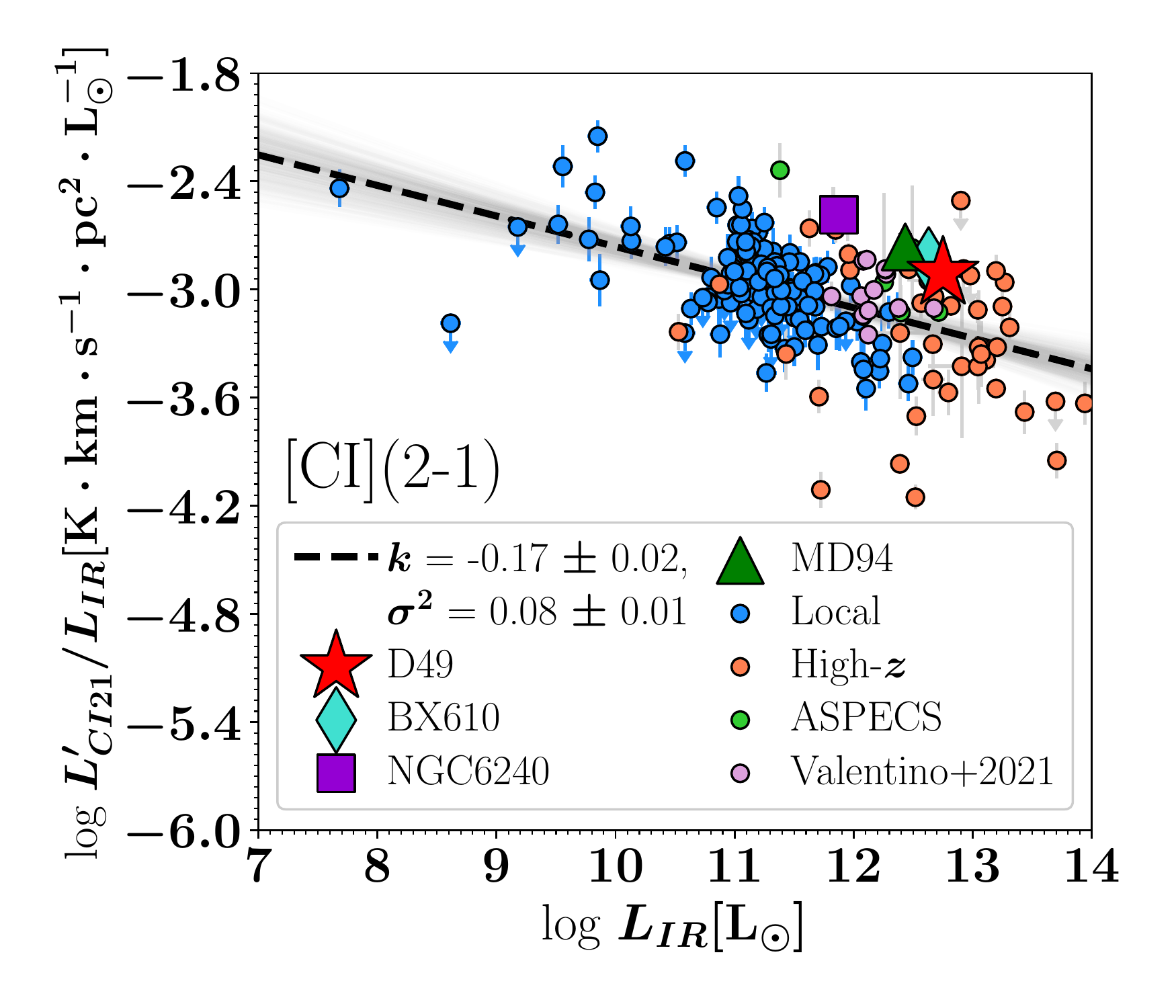}
    \quad
    \includegraphics[width=0.4\textwidth,trim=0.45cm 10 0.88cm 10,clip]{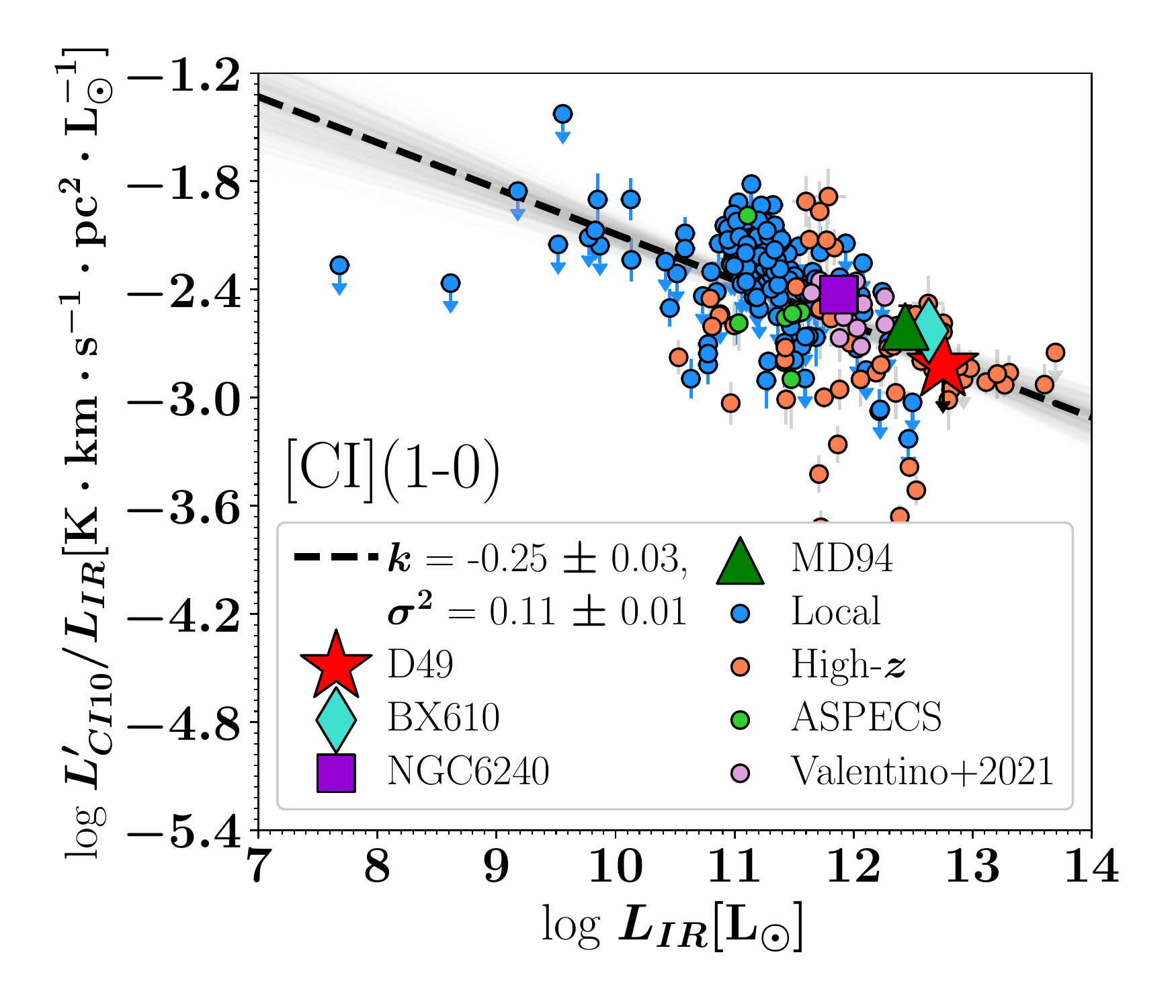}
    \caption{$L'_{\rm line}/L_{\rm IR}$ [\kkmspclsun] ratios as a function of \lir. Clockwise, from  top left: \cofive, \coseven, \cione, and \citwo.  D49 is shown as a red star, BX610 as a turquoise diamond, NGC 6240 as a purple square, and MD94 as a green triangle. The blue and orange points are respectively from the local and high-redshift galaxy samples in \cite{Valentino2020}, green points are the samples from \cite{Boogaard2020}, and pink points are from \cite{Valentino2021}. BX610 and MD94 are not shown in the \cofive-IR plane (\textit{top left}) due to lack of \cofive\, observations. The black dashed and grey lines indicate the best fit in the $\mathrm{log}(L_{\rm IR})$-$\mathrm{log}(L'_{\rm line})$ space and random sampling of the posterior distribution. Two galaxies in \cite{Valentino2021} are found to have high \lprimecofive/\lir\ ratios (pink  down-pointing triangles in the upper left panel), both of which are also found to have strong AGN signatures ($f_{\rm AGN} \sim 0.9$). Three galaxies in \cite{Boogaard2020} are found to have high \lprimecoseven/\lir\ ratios (green  down-pointing triangles in the upper right panel, one of which is obscured by purple square). AGN signatures are detected in two of them ($f_{\rm AGN} \sim 0.08$), while the other one, with the highest \lprimecoseven/\lir\ ratio, is labelled as a non-AGN.  }
    \label{fig.dis1}
\end{figure*}
    
In Fig. \ref{fig.dis1}, we show the location of D49 in the \lir-$(L'_{\rm line}/L_{\rm IR})$ planes.
As a comparison, we plot the local and high-redshift samples in \cite{Valentino2020, Valentino2021} and galaxies from the ALMA Spectroscopic Survey in the Hubble Ultra Deep Field \citep[ASPECS,][]{Boogaard2020}. We recomputed the best-fit linear correlations between line luminosities and \lir\ in the logarithmic space 
using \href{https://github.com/jmeyers314/linmix}{\ttfamily linmix} \citep{Kelly2007}. 
The best-fit parameters and intrinsic scatters ($\sigma \lesssim 0.3\,\mathrm{dex}$) are reported in Fig. \ref{fig.dis1} and are  
consistent with previous works \citep{Greve2014, Liu2015, Daddi2015, Yang2017, Valentino2020}. 
D49 displays some of the highest \lprimecofive\, and $L_{\rm CO(7-6)}^\prime / L_{\rm IR}$ ratios in our literature compilation, \tcb{falling in the 93rd and 96th percentiles} of the distributions. The measured \lprimecofive/\lir\ and \lprimecoseven/\lir\ ratios are $1.9\times$ ($1.3\sigma$) and $3.3\times$ ($2.4\sigma$) higher than the medians of the ensemble of local and distant galaxies. These differences further increase when comparing D49 with \tcb{IR and sub-millimetre selected }high-redshift galaxies only ($2.1\times$ and $3.7\times$ times higher at $1.7\sigma$ and $2.0\sigma$ for $J_{\rm up}=5$ and $7$, respectively). The significance is slightly reduced owing to the smaller number statistics increasing the scatter. Interestingly, the ratios are more consistent with those from the blind ASPECS survey \citep{Boogaard2020}. However, we note that 
$25\%$ ($62.5\%$) of the ASPECS sources with \cofive \,(\coseven) detections in Fig. \ref{fig.dis1}
are X-ray detected sources and candidate AGN -- which might boost the molecular gas excitation.
Our target is also consistent with the literature sample and general trends in the \lir-(\ci/\lir)\ luminosity planes.

\subsection{Potential gas heating mechanisms}

\begin{figure}
   \centering
   \includegraphics[width=\columnwidth,trim={0 0.5cm 0 0.5cm},clip]{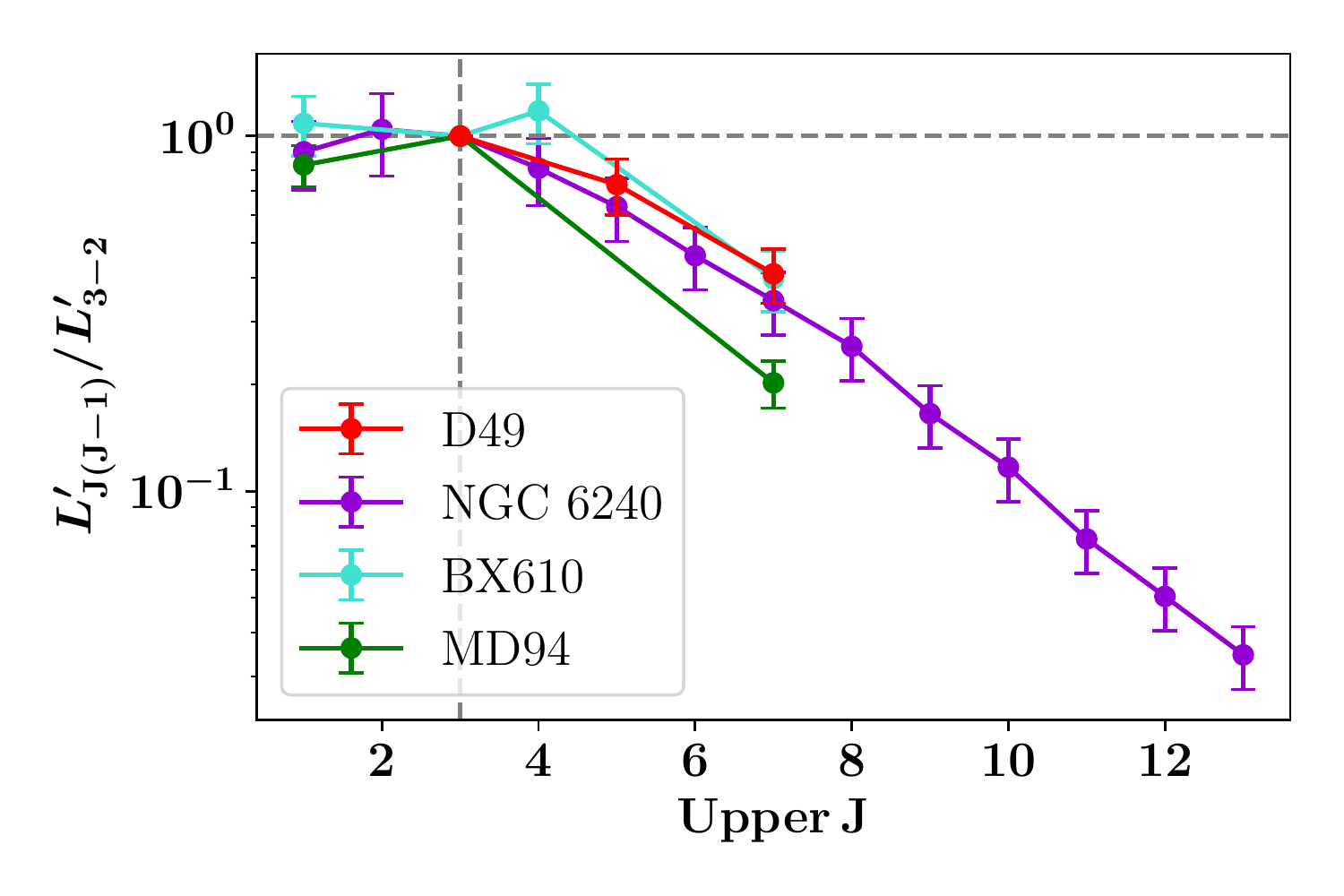}
   \caption{CO spectral line energy distributions of D49, compared to BX610, NGC 6240, and MD94, normalised by their own $L'_{\rm CO(3-2)}$ (see inset for colour-coding).}
   \label{fig.lineratio}
   \end{figure}

In the literature samples mentioned above, we highlighted Q2343-BX610 and Q1700-MD94 (hereafter BX610 and MD94), two massive star-forming main-sequence galaxies at $z \sim 2$ \citep{Brisbin2019, Henriquez2022}. These two galaxies have similar line and continuum coverage to D49, despite being at slightly lower redshifts. We also note NGC 6240, a widely studied local galaxy (e.g. \citealt{Komossa2003, Rieke1985, Vignati1999}). The reason is clear from Fig. \ref{fig.dis1}: all these galaxies occupy similar loci in the $ L'_{\rm line}/L_{\rm IR} $ versus $ L_{\rm IR} $ relations. For D49, we estimate $\log(L_{\rm CO(7-6)}^\prime/L_{\rm IR}\,\mathrm{K\,km\,s^{-1}\,pc^2\,}L_{\odot}^{-1}) = (-2.50 \pm 0.04)$, while this  value is $\sim -2.32$ for NGC 6240 \citep{Meijerink2013}, and $\sim -2.69$ for BX610 \citep{Brisbin2019}. For MD94, we derive its $ L_{\rm IR}\,[L_\odot]$ from the ${\rm SFR} = 271\,M_\odot\,{\rm yr^{-1}}$ \citep{Tacconi2013} with the same conversion adopted above. Also in this case we find $\log(L_{\rm CO(7-6)}^\prime/L_{\rm IR}\,\mathrm{K\,km\,s^{-1}\,pc^2\,}L_{\odot}^{-1}) \simeq -2.50$ consistent with that of D49. All these galaxies exhibit a \coseven\ emission brighter than what is expected from their \lir. Inverting the argument, following Eq.(1) in \cite{Lu2015}, the SFR of D49 derived by $L_{\rm CO(7-6)}$ [\lsun] would be $\sim 3500\, M_\odot/{\rm yr}$, which is $\sim6\times$ larger than what we derive from the SED fitting ($\mathrm{SFR}\sim 600\, M_\odot/{\rm yr}$). Figure \ref{fig.lineratio} presents the CO spectral line energy distributions of these four galaxies, normalised by $L'_{\rm CO(3-2)}$. Despite the undersampling at low-$J$ for D49, its \tcb{spectral line energy distribution (SLED)} shows a similar trend to those of NGC 6240 and BX610 at $J>3$, while looking more excited than MD94.

\cite{Brisbin2019} showed how a single photon-dominated region (PDR) model is not sufficient to explain the line ratios in BX610. \tcb{At least one additional warm PDR component} should be added, considering this heating source only. However, since the CO emission is brighter than   expected from $ L_{\rm IR}$, and thus from SFR-driven heating mechanisms, one would either conclude that either \coseven\ is not a good SFR tracer \citep{Brisbin2019} or that other possibilities not affecting $ L_{\rm IR}$ should be considered.

\cite{Brisbin2019} proposed slow shocks as  alternative contributors to the CO excitation in BX610. These were  brought forward also in the case of NGC 6240 \citep{Meijerink2013, Lu2014}. However, while shocks in NGC 6240 are driven by the collision of two galaxies, their origin in BX610 and other high-redshift galaxies on and above the main sequence remains unknown. Turbulence could also contribute to the heating of the ISM in strongly star-forming galaxies \citep{Harrington2021}. High-redshift galaxies are more gas rich than local objects and the ISM densities and temperatures increase at least up to $z \sim 1.5$ \citep{Liu2021}, which naturally leads to more turbulent ISM and shocks originating from higher Mach numbers.

X-ray dominated regions \citep[XDRs,][]{Wolfire2022} could also contribute to the enhancement of high-$J$ CO transitions. This might be the case for MD94, for which \cite{Erb2006} determined the existence of an AGN from spectra. In addition, signatures of low-luminosity AGN  have  been found in NGC 6240 and BX610 \citep{Rosenberg2015ApJ...801...72R, Bolatto2015ApJ...809..175B}. Interestingly, the CO excitation in MD94 does not seem as high as for the remaining three galaxies, at least up to \coseven\ (Fig. \ref{fig.lineratio}). The absence of X-ray emission \citep{Magdis2017} and strong mid-IR excess in our target D49 makes the AGN solution less favourable.

At this stage, the relatively small number of transitions observed for D49 and other high-redshift objects does not allow us to strongly constrain the contribution of different heating mechanisms. For reference only, large velocity gradient (LVG) modelling with a minimal set of free parameters (gas densities $n_{\rm H}$, kinetic temperatures $T_{\rm kin}$, and [CI]/CO ratios; \citealt{Liu2021}) returns $\mathrm{log}(n_{\rm H}/\mathrm{cm^{-3}})=3.7\pm0.2$, but not meaningful constraints on the remaining variables, \tcb{let alone adding two phases.} A simple PDR model ({\ttfamily pdrtpy}, \citealt{Pound2023}) would favour $\sim1$ dex higher densities, inconsistent with the LVG result, and would suffer from a series of limitations that are well documented in the literature \citep{Papadopoulos2004, Papadopoulos2022, Harrington2021, Liu2021}. Slow C-shocks in high-density environments ($v\sim 10$ \kms, $n_{\rm H}\sim10^4\,\mathrm{cm^{-3}}$) would be able to reproduce some of the CO line ratios observed in D49 (\lprimecoseven$<$\lprimecofive\ and \lprimecothree; Table A1 in \citealt{Flower2010}).

\subsection{Benchmarking gas tracers in a main-sequence galaxy}
\label{sec:gas_estimates}
D49 offers the opportunity to estimate \mgas\ from several different tracers and cross-check their calibrations. We summarise the results of our multi-variate approach in Table \ref{tab.mass} in the  Appendix.\\

First, we re-derived \mgas\ from \cothree\, applying an excitation correction factor of $r_{31}=0.5\pm0.15$ and a CO conversion factor of $\alpha_{\rm CO}=3.5$ consistent with \cite{Magdis2017}. We then turned to dust-based estimates and derived \mgas\ from the full SED modelling with \texttt{Stardust} or single-band extrapolations from the Rayleigh-Jeans tail \citep{Scoville2016}. We allowed for metallicity-dependent gas-to-dust ratios $\delta_{\rm GDR}(Z)$ at fixed solar metallicity or for implementing the FMR (Sect. \ref{sec:modeling}). Finally, we took advantage of the availability of both \ci\ transitions. As local thermodynamic equilibrium (LTE) is not necessarily fulfilled in galaxies \citep{Papadopoulos2022}, we derived hydrogen mass under non-LTE assumption as
\begin{eqnarray}
\label{eq.gasmasslte}
M_{\rm H_2} = 1375.8 D_{\rm L}^2 (1+z)^{-1} \left(\frac{X[CI]}{10^{-5}}\right)^{-1}\left(\frac{A_{10}}{10^{-7}{\rm s^{-1}}}\right)^{-1} \\ \nonumber
\times Q_{10}^{-1}S_{\rm [CI](1-0)}\Delta v,
\end{eqnarray}
where $A_{10} = 10^{-7.10}\,{\rm s^{-1}}$ is the Einstein A coefficient \citep{Papadopoulos2004, Bothwell2017}. We adopt the neutral carbon abundance for main-sequence galaxies at $z\sim1.2$ in \cite{Valentino2018} $\log X[CI] = (-4.7 \pm 0.1)$. 
According to our measurements, we find a \citwos/\ciones\, line ratio of $\mathcal{R} = S_{\rm [CI](2-1)}\Delta v / S_{\rm [CI](1-0)}\Delta v > 2.25 $, where $S\Delta v$ are the velocity integrated fluxes in \jykms. We then derived the  kinetic temperature of [CI] as $T_{\rm kin} = \alpha T_{\rm dust}$ adopting $\alpha = 1.2$ for main-sequence galaxies \citep{Papadopoulos2022} and dust temperature $T_{\rm dust} = 41\,\mathrm{K}$ \citep{Magdis2017}. Thus, $T_{\rm kin} \sim 49\,{\rm K}$. For a typical gas density of $ n \approx 10^{4}\,{\rm cm^{-3}}$, we expect excitation corrections $Q_{10} \approx 0.45$ and $Q_{21} \approx 0.3$ \citep[Figures 4 and 5 in][]{Papadopoulos2022}.
We note that the helium contribution is not included in Eq. \ref{eq.gasmasslte}. We thus include an additional factor of 1.36 to account for the latter, thus resulting in a final estimate of $\log (M_{\rm gas}/M_\odot) < 11.22 $ from \ciones. By replacing all the \ciones-related terms with those for \citwos, we obtain a consistent \mgas\ upper limit,  set by the uncertainty on \ciones. 
As a cross-check, we applied the recent calibration by \cite{Dunne2022MNRAS.tmp.1998D}. The authors found that \ciones\, is the preferred tracer for galaxies with $\log (L_{\rm IR}/L_\odot) > 11$ with the mean conversion factor $\alpha_{\rm CI} = (17.0 \pm 0.3 \, M_\odot)\, \mathrm{[K\,km\,s^{-1}\,pc^2]^{-1}} $, based on a sample of 407 metal-rich galaxies ranging from local to $ z \approx 6 $. Given our upper limit on $L'_{\rm [CI](1-0)}$, we derive $M_{\rm gas} = \alpha_{\rm CI}L'_{\rm [CI](1-0)} < 10^{11.16}\,M_\odot$, consistent with our estimates. As noted in \cite{Dunne2022MNRAS.tmp.1998D}, caution should be exerted when deriving \mgas\ from \citwos\ given the uncertain excitation conditions.\\

All the \mgas\ estimates for D49 are consistent within the uncertainties and a factor of three or less, the highest being that from \cothree. However, given the high CO excitation, a higher $r_{31}$ value for D49 could be expected. Among the examples detailed in the previous section, $r_{31} = 1.1\pm0.2$ is found for NGC 6240, $0.9 \pm 0.2$ for BX610, and $1.2 \pm 0.2$ for MD94. By adopting the average of these values $\left\langle r_{31} \right\rangle = 1.1 \pm 0.1 $ (consistent with thermalisation up to $J=3$), we find $\log(M_{\rm gas}/M_\odot) = 11.14 \pm 0.08$, more in line with the rest of the values in Table \ref{tab.mass}.

\subsection{Massive galaxy at the last stage of its evolution}
Even considering a conservative upper limit of $M_{\rm gas} < 10^{11.22} M_\odot$, we estimate a gas depletion timescale of $ \tau_{\rm dep} = M_{\rm gas} / {\rm SFR} < 0.29 \, {\rm Gyr}$. For reference, for their main-sequence sample at $z = 3.2$, \cite{Schinnerer2016} found an average $\tau_{\rm dep} = 0.68^{+0.07}_{-0.08} \, {\rm Gyr}$. The high SFE of D49 is typical of starburst galaxies, but its location is on the main sequence (the distance from the parametrisation of \citealt{Sargent2014} is $\Delta \mathrm{MS} = 
 \mathrm{SFR} / \mathrm{SFR}_{\rm MS} = 0.9$). 
Therefore, if the gas reservoirs are not replenished, D49 will exhaust its gas mass in a relatively short time, and will possibly transit to quiescence. This would be consistent with the fact that D49 is a very massive object in the bending part of the main sequence \citep{Schreiber2015}. \tcb{We might thus be  witnessing the ending growth spurt of this massive galaxy. }

\section{Conclusions}

We presented robust ($\gtrsim10\sigma$) NOEMA measurements of \cofive, \coseven, \citwos, and a conservative upper limit on the \ciones\, emission from the massive main-sequence galaxy D49 at $z = 2.846$. These add to previously available \cothree\ measurements from the literature. Armed with these five lines and exquisite optical to millimetre photometry, we find the following:
   \begin{enumerate}
      \item We measure a higher  \coseven-to-IR ratio than the average at low and high redshifts.
      With such a ratio, the SFR predicted by \coseven\, according to empirical \lir-\lcoseven\, relations would be approximately six times larger than that derived from its infrared luminosity. 
      Either \coseven\ is not a good tracer of SFR in D49 or mechanisms enhancing the high-$J$ CO emission, but not affecting $L_{\rm IR}$, should be considered. 
      \item By comparing D49 with selected IR-bright galaxies in the local universe (NGC 6240) or objects on the main sequence at $z \sim 2$ (BX610, MD94) displaying similar line ratios and \coseven\ enhancements, we consider shocks or mechanical heating in general as a possible cause of the excess \coseven\, emission. 
      The physical drivers of mechanical heating are unknown for D49, but, as noted in the past, the high gas densities and temperatures are conditions naturally favouring turbulence, of which high CO excitation could be a reflection.
      \item We benchmark gas mass estimates from CO, \ci, and dust under several assumptions in D49, finding them overall consistent within a factor of $<3\times$. 
     \item Given its high $\mathrm{SFR}\sim600$ \myr, the ensuing gas depletion time of D49 is $<0.3$ Gyr, indicating that it might possibly transition to quiescence shortly if no replenishment is ensured.     
   \end{enumerate}

Future observations mapping more CO lines would help us to determine the SLED at much higher accuracy, also anchoring to low-$J$ transitions. Direct observations of $\mathrm{H}_2$ lines (as a shock tracer) in the mid-IR with \jwst\ would allow us to constrain the conversion factor and understand what contributes to the mechanical heating in D49.

\begin{acknowledgements}
The authors warmly thank the anonymous referee for the insightful comments that improved the presentation of our work and Isabella Cortzen for her contribution to the data reduction. The Cosmic Dawn Center (DAWN) is funded by the Danish National Research Foundation under grant No. 140. SJ is supported by the European Union's Horizon Europe research and innovation program under the Marie Sk\l{}odowska-Curie grant agreement No. 101060888.
\end{acknowledgements}

\bibliographystyle{aa}
\bibliography{paper_ref}

\begin{thebibliography}{79}
\expandafter\ifx\csname natexlab\endcsname\relax\def\natexlab#1{#1}\fi

\bibitem[{{Battisti} {et~al.}(2019){Battisti}, {da Cunha}, {Grasha}, {Salvato},
  {Daddi}, {Davies}, {Jin}, {Liu}, {Schinnerer}, {Vaccari}, \& {COSMOS
  Collaboration}}]{Battisti2019}
{Battisti}, A.~J., {da Cunha}, E., {Grasha}, K., {et~al.} 2019, \apj, 882, 61

\bibitem[{{Berta} {et~al.}(2016){Berta}, {Lutz}, {Genzel},
  {F{\"o}rster-Schreiber}, \& {Tacconi}}]{Berta2016}
{Berta}, S., {Lutz}, D., {Genzel}, R., {F{\"o}rster-Schreiber}, N.~M., \&
  {Tacconi}, L.~J. 2016, \aap, 587, A73

\bibitem[{{Bolatto} {et~al.}(2015){Bolatto}, {Warren}, {Leroy}, {Tacconi},
  {Bouch{\'e}}, {F{\"o}rster Schreiber}, {Genzel}, {Cooper}, {Fisher},
  {Combes}, {Garc{\'\i}a-Burillo}, {Burkert}, {Bournaud}, {Weiss}, {Saintonge},
  {Wuyts}, \& {Sternberg}}]{Bolatto2015ApJ...809..175B}
{Bolatto}, A.~D., {Warren}, S.~R., {Leroy}, A.~K., {et~al.} 2015, \apj, 809,
  175

\bibitem[{{Boogaard} {et~al.}(2020){Boogaard}, {van der Werf}, {Weiss},
  {Popping}, {Decarli}, {Walter}, {Aravena}, {Bouwens}, {Riechers},
  {Gonz{\'a}lez-L{\'o}pez}, {Smail}, {Carilli}, {Kaasinen}, {Daddi}, {Cox},
  {D{\'\i}az-Santos}, {Inami}, {Cortes}, \& {Wagg}}]{Boogaard2020}
{Boogaard}, L.~A., {van der Werf}, P., {Weiss}, A., {et~al.} 2020, \apj, 902,
  109

\bibitem[{{Boquien} {et~al.}(2019){Boquien}, {Burgarella}, {Roehlly}, {Buat},
  {Ciesla}, {Corre}, {Inoue}, \& {Salas}}]{Boquien2019}
{Boquien}, M., {Burgarella}, D., {Roehlly}, Y., {et~al.} 2019, \aap, 622, A103

\bibitem[{{Bothwell} {et~al.}(2017){Bothwell}, {Aguirre}, {Aravena},
  {Bethermin}, {Bisbas}, {Chapman}, {De Breuck}, {Gonzalez}, {Greve},
  {Hezaveh}, {Ma}, {Malkan}, {Marrone}, {Murphy}, {Spilker}, {Strandet},
  {Vieira}, \& {Wei{\ss}}}]{Bothwell2017}
{Bothwell}, M.~S., {Aguirre}, J.~E., {Aravena}, M., {et~al.} 2017, \mnras, 466,
  2825

\bibitem[{{Bothwell} {et~al.}(2013){Bothwell}, {Smail}, {Chapman}, {Genzel},
  {Ivison}, {Tacconi}, {Alaghband-Zadeh}, {Bertoldi}, {Blain}, {Casey}, {Cox},
  {Greve}, {Lutz}, {Neri}, {Omont}, \& {Swinbank}}]{Bothwell2013}
{Bothwell}, M.~S., {Smail}, I., {Chapman}, S.~C., {et~al.} 2013, \mnras, 429,
  3047

\bibitem[{{Bourne} {et~al.}(2019){Bourne}, {Dunlop}, {Simpson}, {Rowlands},
  {Geach}, \& {McLeod}}]{Bourne2019}
{Bourne}, N., {Dunlop}, J.~S., {Simpson}, J.~M., {et~al.} 2019, \mnras, 482,
  3135

\bibitem[{{Brammer} {et~al.}(2008){Brammer}, {van Dokkum}, \&
  {Coppi}}]{Brammer2008}
{Brammer}, G.~B., {van Dokkum}, P.~G., \& {Coppi}, P. 2008, \apj, 686, 1503

\bibitem[{{Brisbin} {et~al.}(2019){Brisbin}, {Aravena}, {Daddi}, {Dannerbauer},
  {Decarli}, {Gonz{\'a}lez-L{\'o}pez}, {Riechers}, \& {Wagg}}]{Brisbin2019}
{Brisbin}, D., {Aravena}, M., {Daddi}, E., {et~al.} 2019, \aap, 628, A104

\bibitem[{{Burgarella} {et~al.}(2005){Burgarella}, {Buat}, \&
  {Iglesias-P{\'a}ramo}}]{Burgarella2005}
{Burgarella}, D., {Buat}, V., \& {Iglesias-P{\'a}ramo}, J. 2005, \mnras, 360,
  1413

\bibitem[{{Calistro Rivera} {et~al.}(2018){Calistro Rivera}, {Hodge}, {Smail},
  {Swinbank}, {Weiss}, {Wardlow}, {Walter}, {Rybak}, {Chen}, {Brandt},
  {Coppin}, {da Cunha}, {Dannerbauer}, {Greve}, {Karim}, {Knudsen},
  {Schinnerer}, {Simpson}, {Venemans}, \& {van der
  Werf}}]{Gabriela2018ApJ...863...56C}
{Calistro Rivera}, G., {Hodge}, J.~A., {Smail}, I., {et~al.} 2018, \apj, 863,
  56

\bibitem[{{Carilli} \& {Walter}(2013)}]{Carilli2013}
{Carilli}, C.~L. \& {Walter}, F. 2013, \araa, 51, 105

\bibitem[{{Chabrier}(2003)}]{Chabrier2003}
{Chabrier}, G. 2003, \pasp, 115, 763

\bibitem[{{Chen} {et~al.}(2017){Chen}, {Hodge}, {Smail}, {Swinbank}, {Walter},
  {Simpson}, {Calistro Rivera}, {Bertoldi}, {Brandt}, {Chapman}, {da Cunha},
  {Dannerbauer}, {De Breuck}, {Harrison}, {Ivison}, {Karim}, {Knudsen},
  {Wardlow}, {Wei{\ss}}, \& {van der Werf}}]{Chen2017ApJ...846..108C}
{Chen}, C.-C., {Hodge}, J.~A., {Smail}, I., {et~al.} 2017, \apj, 846, 108

\bibitem[{{Cochrane} {et~al.}(2021){Cochrane}, {Best}, {Smail}, {Ibar},
  {Cheng}, {Swinbank}, {Molina}, {Sobral}, \&
  {Dudzevi{\v{c}}i{\={u}}t{\.{e}}}}]{Cochrane2021MNRAS.503.2622C}
{Cochrane}, R.~K., {Best}, P.~N., {Smail}, I., {et~al.} 2021, \mnras, 503, 2622

\bibitem[{{da Cunha} {et~al.}(2008){da Cunha}, {Charlot}, \&
  {Elbaz}}]{deCunha2008}
{da Cunha}, E., {Charlot}, S., \& {Elbaz}, D. 2008, \mnras, 388, 1595

\bibitem[{{Daddi} {et~al.}(2015){Daddi}, {Dannerbauer}, {Liu}, {Aravena},
  {Bournaud}, {Walter}, {Riechers}, {Magdis}, {Sargent}, {B{\'e}thermin},
  {Carilli}, {Cibinel}, {Dickinson}, {Elbaz}, {Gao}, {Gobat}, {Hodge}, \&
  {Krips}}]{Daddi2015}
{Daddi}, E., {Dannerbauer}, H., {Liu}, D., {et~al.} 2015, \aap, 577, A46

\bibitem[{{Daddi} {et~al.}(2007){Daddi}, {Dickinson}, {Morrison}, {Chary},
  {Cimatti}, {Elbaz}, {Frayer}, {Renzini}, {Pope}, {Alexander}, {Bauer},
  {Giavalisco}, {Huynh}, {Kurk}, \& {Mignoli}}]{Daddi2007}
{Daddi}, E., {Dickinson}, M., {Morrison}, G., {et~al.} 2007, \apj, 670, 156

\bibitem[{{Daddi} {et~al.}(2010){Daddi}, {Elbaz}, {Walter}, {Bournaud},
  {Salmi}, {Carilli}, {Dannerbauer}, {Dickinson}, {Monaco}, \&
  {Riechers}}]{Daddi2010}
{Daddi}, E., {Elbaz}, D., {Walter}, F., {et~al.} 2010, \apjl, 714, L118

\bibitem[{{Delvecchio} {et~al.}(2021){Delvecchio}, {Daddi}, {Sargent},
  {Jarvis}, {Elbaz}, {Jin}, {Liu}, {Whittam}, {Algera}, {Carraro}, {D'Eugenio},
  {Delhaize}, {Kalita}, {Leslie}, {Moln{\'a}r}, {Novak}, {Prandoni},
  {Smol{\v{c}}i{\'c}}, {Ao}, {Aravena}, {Bournaud}, {Collier},
  {Randriamampandry}, {Randriamanakoto}, {Rodighiero}, {Schober}, {White}, \&
  {Zamorani}}]{Delvecchio2021}
{Delvecchio}, I., {Daddi}, E., {Sargent}, M.~T., {et~al.} 2021, \aap, 647, A123

\bibitem[{{Draine} {et~al.}(2014){Draine}, {Aniano}, {Krause}, {Groves},
  {Sandstrom}, {Braun}, {Leroy}, {Klaas}, {Linz}, {Rix}, {Schinnerer},
  {Schmiedeke}, \& {Walter}}]{Draine2014}
{Draine}, B.~T., {Aniano}, G., {Krause}, O., {et~al.} 2014, \apj, 780, 172

\bibitem[{{Draine} \& {Li}(2007)}]{DL07}
{Draine}, B.~T. \& {Li}, A. 2007, \apj, 657, 810

\bibitem[{{Dunne} {et~al.}(2022){Dunne}, {Maddox}, {Papadopoulos}, {Ivison}, \&
  {Gomez}}]{Dunne2022MNRAS.tmp.1998D}
{Dunne}, L., {Maddox}, S.~J., {Papadopoulos}, P.~P., {Ivison}, R.~J., \&
  {Gomez}, H.~L. 2022, \mnras [\eprint[arXiv]{2208.01622}]

\bibitem[{{Elbaz} {et~al.}(2007){Elbaz}, {Daddi}, {Le Borgne}, {Dickinson},
  {Alexander}, {Chary}, {Starck}, {Brandt}, {Kitzbichler}, {MacDonald},
  {Nonino}, {Popesso}, {Stern}, \& {Vanzella}}]{Elbaz2007}
{Elbaz}, D., {Daddi}, E., {Le Borgne}, D., {et~al.} 2007, \aap, 468, 33

\bibitem[{{Erb} {et~al.}(2006){Erb}, {Steidel}, {Shapley}, {Pettini}, {Reddy},
  \& {Adelberger}}]{Erb2006}
{Erb}, D.~K., {Steidel}, C.~C., {Shapley}, A.~E., {et~al.} 2006, \apj, 647, 128

\bibitem[{{Flower} \& {Pineau Des For{\^e}ts}(2010)}]{Flower2010}
{Flower}, D.~R. \& {Pineau Des For{\^e}ts}, G. 2010, \mnras, 406, 1745

\bibitem[{{Genzel} {et~al.}(2010){Genzel}, {Tacconi}, {Gracia-Carpio},
  {Sternberg}, {Cooper}, {Shapiro}, {Bolatto}, {Bouch{\'e}}, {Bournaud},
  {Burkert}, {Combes}, {Comerford}, {Cox}, {Davis}, {F{\"o}rster Schreiber},
  {Garcia-Burillo}, {Lutz}, {Naab}, {Neri}, {Omont}, {Shapley}, \&
  {Weiner}}]{Genzel2010mn}
{Genzel}, R., {Tacconi}, L.~J., {Gracia-Carpio}, J., {et~al.} 2010, \mnras,
  407, 2091

\bibitem[{{Greve} {et~al.}(2014){Greve}, {Leonidaki}, {Xilouris}, {Wei{\ss}},
  {Zhang}, {van der Werf}, {Aalto}, {Armus}, {D{\'\i}az-Santos}, {Evans},
  {Fischer}, {Gao}, {Gonz{\'a}lez-Alfonso}, {Harris}, {Henkel}, {Meijerink},
  {Naylor}, {Smith}, {Spaans}, {Stacey}, {Veilleux}, \& {Walter}}]{Greve2014}
{Greve}, T.~R., {Leonidaki}, I., {Xilouris}, E.~M., {et~al.} 2014, \apj, 794,
  142

\bibitem[{{Harrington} {et~al.}(2021){Harrington}, {Weiss}, {Yun}, {Magnelli},
  {Sharon}, {Leung}, {Vishwas}, {Wang}, {Frayer}, {Jim{\'e}nez-Andrade}, {Liu},
  {Garc{\'\i}a}, {Romano-D{\'\i}az}, {Frye}, {Jarugula}, {B{\u{a}}descu},
  {Berman}, {Dannerbauer}, {D{\'\i}az-S{\'a}nchez}, {Grassitelli},
  {Kamieneski}, {Kim}, {Kirkpatrick}, {Lowenthal}, {Messias}, {Puschnig},
  {Stacey}, {Torne}, \& {Bertoldi}}]{Harrington2021}
{Harrington}, K.~C., {Weiss}, A., {Yun}, M.~S., {et~al.} 2021, \apj, 908, 95

\bibitem[{{Henr{\'\i}quez-Brocal} {et~al.}(2022){Henr{\'\i}quez-Brocal},
  {Herrera-Camus}, {Tacconi}, {Genzel}, {Bolatto}, {Bovino}, {Demarco},
  {F{\"o}rster Schreiber}, {Lee}, {Lutz}, \& {Rubio}}]{Henriquez2022}
{Henr{\'\i}quez-Brocal}, K., {Herrera-Camus}, R., {Tacconi}, L., {et~al.} 2022,
  \aap, 657, L15

\bibitem[{{H{\"o}gbom}(1974)}]{Hogbom1974}
{H{\"o}gbom}, J.~A. 1974, \aaps, 15, 417

\bibitem[{{Kaasinen} {et~al.}(2019){Kaasinen}, {Scoville}, {Walter}, {Da
  Cunha}, {Popping}, {Pavesi}, {Darvish}, {Casey}, {Riechers}, \&
  {Glover}}]{Kaasinen2019}
{Kaasinen}, M., {Scoville}, N., {Walter}, F., {et~al.} 2019, \apj, 880, 15

\bibitem[{{Kelly}(2007)}]{Kelly2007}
{Kelly}, B.~C. 2007, \apj, 665, 1489

\bibitem[{{Kennicutt}(1998{\natexlab{a}})}]{Kennicutt1998rev}
{Kennicutt}, Robert~C., J. 1998{\natexlab{a}}, \araa, 36, 189

\bibitem[{{Kennicutt}(1998{\natexlab{b}})}]{Kennicutt1998}
{Kennicutt}, Robert~C., J. 1998{\natexlab{b}}, \apj, 498, 541

\bibitem[{{Kokorev} {et~al.}(2021){Kokorev}, {Magdis}, {Davidzon}, {Brammer},
  {Valentino}, {Daddi}, {Ciesla}, {Liu}, {Jin}, {Cortzen}, {Delvecchio},
  {Gim{\'e}nez-Arteaga}, {G{\'o}mez-Guijarro}, {Sargent}, {Toft}, \&
  {Weaver}}]{Kokorev2021}
{Kokorev}, V.~I., {Magdis}, G.~E., {Davidzon}, I., {et~al.} 2021, \apj, 921, 40

\bibitem[{{Komossa} {et~al.}(2003){Komossa}, {Burwitz}, {Hasinger}, {Predehl},
  {Kaastra}, \& {Ikebe}}]{Komossa2003}
{Komossa}, S., {Burwitz}, V., {Hasinger}, G., {et~al.} 2003, \apjl, 582, L15

\bibitem[{{Liu} {et~al.}(2021){Liu}, {Daddi}, {Schinnerer}, {Saito}, {Leroy},
  {Silverman}, {Valentino}, {Magdis}, {Gao}, {Jin}, {Puglisi}, \&
  {Groves}}]{Liu2021}
{Liu}, D., {Daddi}, E., {Schinnerer}, E., {et~al.} 2021, \apj, 909, 56

\bibitem[{{Liu} {et~al.}(2015){Liu}, {Gao}, {Isaak}, {Daddi}, {Yang}, {Lu}, \&
  {van der Werf}}]{Liu2015}
{Liu}, D., {Gao}, Y., {Isaak}, K., {et~al.} 2015, \apjl, 810, L14

\bibitem[{{Liu} {et~al.}(2019{\natexlab{a}}){Liu}, {Lang}, {Magnelli},
  {Schinnerer}, {Leslie}, {Fudamoto}, {Bondi}, {Groves}, {Jim{\'e}nez-Andrade},
  {Harrington}, {Karim}, {Oesch}, {Sargent}, {Vardoulaki}, {B{\v{a}}descu},
  {Moser}, {Bertoldi}, {Battisti}, {da Cunha}, {Zavala}, {Vaccari}, {Davidzon},
  {Riechers}, \& {Aravena}}]{Liu2019_supplementary}
{Liu}, D., {Lang}, P., {Magnelli}, B., {et~al.} 2019{\natexlab{a}}, \apjs, 244,
  40

\bibitem[{{Liu} {et~al.}(2019{\natexlab{b}}){Liu}, {Schinnerer}, {Groves},
  {Magnelli}, {Lang}, {Leslie}, {Jim{\'e}nez-Andrade}, {Riechers}, {Popping},
  {Magdis}, {Daddi}, {Sargent}, {Gao}, {Fudamoto}, {Oesch}, \&
  {Bertoldi}}]{Liu2019_letter}
{Liu}, D., {Schinnerer}, E., {Groves}, B., {et~al.} 2019{\natexlab{b}}, \apj,
  887, 235

\bibitem[{{Lu} {et~al.}(2014){Lu}, {Zhao}, {Xu}, {Gao}, {Armus}, {Mazzarella},
  {Isaak}, {Petric}, {Charmandaris}, {D{\'\i}az-Santos}, {Evans}, {Howell},
  {Appleton}, {Inami}, {Iwasawa}, {Leech}, {Lord}, {Sanders}, {Schulz},
  {Surace}, \& {van der Werf}}]{Lu2014}
{Lu}, N., {Zhao}, Y., {Xu}, C.~K., {et~al.} 2014, \apjl, 787, L23

\bibitem[{{Lu} {et~al.}(2015){Lu}, {Zhao}, {Xu}, {Gao}, {D{\'\i}az-Santos},
  {Charmandaris}, {Inami}, {Howell}, {Liu}, {Armus}, {Mazzarella}, {Privon},
  {Lord}, {Sanders}, {Schulz}, \& {van der Werf}}]{Lu2015}
{Lu}, N., {Zhao}, Y., {Xu}, C.~K., {et~al.} 2015, \apjl, 802, L11

\bibitem[{{Madau} \& {Dickinson}(2014)}]{MadauDickinson2014}
{Madau}, P. \& {Dickinson}, M. 2014, \araa, 52, 415

\bibitem[{{Madden} {et~al.}(2020){Madden}, {Cormier}, {Hony}, {Lebouteiller},
  {Abel}, {Galametz}, {De Looze}, {Chevance}, {Polles}, {Lee}, {Galliano},
  {Lambert-Huyghe}, {Hu}, \& {Ramambason}}]{Madden2020}
{Madden}, S.~C., {Cormier}, D., {Hony}, S., {et~al.} 2020, \aap, 643, A141

\bibitem[{{Magdis} {et~al.}(2012){Magdis}, {Daddi}, {B{\'e}thermin}, {Sargent},
  {Elbaz}, {Pannella}, {Dickinson}, {Dannerbauer}, {da Cunha}, {Walter},
  {Rigopoulou}, {Charmandaris}, {Hwang}, \& {Kartaltepe}}]{Magdis2012}
{Magdis}, G.~E., {Daddi}, E., {B{\'e}thermin}, M., {et~al.} 2012, \apj, 760, 6

\bibitem[{{Magdis} {et~al.}(2011){Magdis}, {Daddi}, {Elbaz}, {Sargent},
  {Dickinson}, {Dannerbauer}, {Aussel}, {Walter}, {Hwang}, {Charmandaris},
  {Hodge}, {Riechers}, {Rigopoulou}, {Carilli}, {Pannella}, {Mullaney},
  {Leiton}, \& {Scott}}]{Magdis2011}
{Magdis}, G.~E., {Daddi}, E., {Elbaz}, D., {et~al.} 2011, \apjl, 740, L15

\bibitem[{{Magdis} {et~al.}(2017){Magdis}, {Rigopoulou}, {Daddi}, {Bethermin},
  {Feruglio}, {Sargent}, {Dannerbauer}, {Dickinson}, {Elbaz}, {Gomez Guijarro},
  {Huang}, {Toft}, \& {Valentino}}]{Magdis2017}
{Magdis}, G.~E., {Rigopoulou}, D., {Daddi}, E., {et~al.} 2017, \aap, 603, A93

\bibitem[{{Mannucci} {et~al.}(2010){Mannucci}, {Cresci}, {Maiolino}, {Marconi},
  \& {Gnerucci}}]{Mannucci2010}
{Mannucci}, F., {Cresci}, G., {Maiolino}, R., {Marconi}, A., \& {Gnerucci}, A.
  2010, \mnras, 408, 2115

\bibitem[{{Meijerink} {et~al.}(2013){Meijerink}, {Kristensen}, {Wei{\ss}}, {van
  der Werf}, {Walter}, {Spaans}, {Loenen}, {Fischer}, {Israel}, {Isaak},
  {Papadopoulos}, {Aalto}, {Armus}, {Charmandaris}, {Dasyra}, {Diaz-Santos},
  {Evans}, {Gao}, {Gonz{\'a}lez-Alfonso}, {G{\"u}sten}, {Henkel}, {Kramer},
  {Lord}, {Mart{\'\i}n-Pintado}, {Naylor}, {Sanders}, {Smith}, {Spinoglio},
  {Stacey}, {Veilleux}, \& {Wiedner}}]{Meijerink2013}
{Meijerink}, R., {Kristensen}, L.~E., {Wei{\ss}}, A., {et~al.} 2013, \apjl,
  762, L16

\bibitem[{{Mullaney} {et~al.}(2011){Mullaney}, {Alexander}, {Goulding}, \&
  {Hickox}}]{Mullaney2011}
{Mullaney}, J.~R., {Alexander}, D.~M., {Goulding}, A.~D., \& {Hickox}, R.~C.
  2011, \mnras, 414, 1082

\bibitem[{{Noeske} {et~al.}(2007){Noeske}, {Weiner}, {Faber}, {Papovich},
  {Koo}, {Somerville}, {Bundy}, {Conselice}, {Newman}, {Schiminovich}, {Le
  Floc'h}, {Coil}, {Rieke}, {Lotz}, {Primack}, {Barmby}, {Cooper}, {Davis},
  {Ellis}, {Fazio}, {Guhathakurta}, {Huang}, {Kassin}, {Martin}, {Phillips},
  {Rich}, {Small}, {Willmer}, \& {Wilson}}]{Noeske2007}
{Noeske}, K.~G., {Weiner}, B.~J., {Faber}, S.~M., {et~al.} 2007, \apjl, 660,
  L43

\bibitem[{{Noll} {et~al.}(2009){Noll}, {Burgarella}, {Giovannoli}, {Buat},
  {Marcillac}, \& {Mu{\~n}oz-Mateos}}]{Noll2009}
{Noll}, S., {Burgarella}, D., {Giovannoli}, E., {et~al.} 2009, \aap, 507, 1793

\bibitem[{{Papadopoulos} {et~al.}(2022){Papadopoulos}, {Dunne}, \&
  {Maddox}}]{Papadopoulos2022}
{Papadopoulos}, P., {Dunne}, L., \& {Maddox}, S. 2022, \mnras, 510, 725

\bibitem[{{Papadopoulos} \& {Greve}(2004)}]{Papadopoulos2004}
{Papadopoulos}, P.~P. \& {Greve}, T.~R. 2004, \apjl, 615, L29

\bibitem[{{Papadopoulos} {et~al.}(2004){Papadopoulos}, {Thi}, \&
  {Viti}}]{Papadodoulos2004b}
{Papadopoulos}, P.~P., {Thi}, W.~F., \& {Viti}, S. 2004, \mnras, 351, 147

\bibitem[{{Pound} \& {Wolfire}(2023)}]{Pound2023}
{Pound}, M.~W. \& {Wolfire}, M.~G. 2023, \aj, 165, 25

\bibitem[{{Rieke} {et~al.}(1985){Rieke}, {Cutri}, {Black}, {Kailey}, {McAlary},
  {Lebofsky}, \& {Elston}}]{Rieke1985}
{Rieke}, G.~H., {Cutri}, R.~M., {Black}, J.~H., {et~al.} 1985, \apj, 290, 116

\bibitem[{{Rigopoulou} {et~al.}(2006){Rigopoulou}, {Huang}, {Papovich},
  {Ashby}, {Barmby}, {Shu}, {Bundy}, {Egami}, {Magdis}, {Smith}, {Willner},
  {Wilson}, \& {Fazio}}]{Rigopoulou2006}
{Rigopoulou}, D., {Huang}, J.~S., {Papovich}, C., {et~al.} 2006, \apj, 648, 81

\bibitem[{{Rodighiero} {et~al.}(2011){Rodighiero}, {Daddi}, {Baronchelli},
  {Cimatti}, {Renzini}, {Aussel}, {Popesso}, {Lutz}, {Andreani}, {Berta},
  {Cava}, {Elbaz}, {Feltre}, {Fontana}, {F{\"o}rster Schreiber},
  {Franceschini}, {Genzel}, {Grazian}, {Gruppioni}, {Ilbert}, {Le Floch},
  {Magdis}, {Magliocchetti}, {Magnelli}, {Maiolino}, {McCracken}, {Nordon},
  {Poglitsch}, {Santini}, {Pozzi}, {Riguccini}, {Tacconi}, {Wuyts}, \&
  {Zamorani}}]{Rodighiero2011}
{Rodighiero}, G., {Daddi}, E., {Baronchelli}, I., {et~al.} 2011, \apjl, 739,
  L40

\bibitem[{{Rodighiero} {et~al.}(2014){Rodighiero}, {Renzini}, {Daddi},
  {Baronchelli}, {Berta}, {Cresci}, {Franceschini}, {Gruppioni}, {Lutz},
  {Mancini}, {Santini}, {Zamorani}, {Silverman}, {Kashino}, {Andreani},
  {Cimatti}, {S{\'a}nchez}, {Le Floch}, {Magnelli}, {Popesso}, \&
  {Pozzi}}]{Rodighiero2014}
{Rodighiero}, G., {Renzini}, A., {Daddi}, E., {et~al.} 2014, \mnras, 443, 19

\bibitem[{{Rosenberg} {et~al.}(2015){Rosenberg}, {van der Werf}, {Aalto},
  {Armus}, {Charmandaris}, {D{\'\i}az-Santos}, {Evans}, {Fischer}, {Gao},
  {Gonz{\'a}lez-Alfonso}, {Greve}, {Harris}, {Henkel}, {Israel}, {Isaak},
  {Kramer}, {Meijerink}, {Naylor}, {Sanders}, {Smith}, {Spaans}, {Spinoglio},
  {Stacey}, {Veenendaal}, {Veilleux}, {Walter}, {Wei{\ss}}, {Wiedner}, {van der
  Wiel}, \& {Xilouris}}]{Rosenberg2015ApJ...801...72R}
{Rosenberg}, M.~J.~F., {van der Werf}, P.~P., {Aalto}, S., {et~al.} 2015, \apj,
  801, 72

\bibitem[{{Sargent} {et~al.}(2014){Sargent}, {Daddi}, {B{\'e}thermin},
  {Aussel}, {Magdis}, {Hwang}, {Juneau}, {Elbaz}, \& {da Cunha}}]{Sargent2014}
{Sargent}, M.~T., {Daddi}, E., {B{\'e}thermin}, M., {et~al.} 2014, \apj, 793,
  19

\bibitem[{{Schinnerer} {et~al.}(2016){Schinnerer}, {Groves}, {Sargent},
  {Karim}, {Oesch}, {Magnelli}, {LeFevre}, {Tasca}, {Civano}, {Cassata}, \&
  {Smol{\v{c}}i{\'c}}}]{Schinnerer2016}
{Schinnerer}, E., {Groves}, B., {Sargent}, M.~T., {et~al.} 2016, \apj, 833, 112

\bibitem[{{Schreiber} {et~al.}(2015){Schreiber}, {Pannella}, {Elbaz},
  {B{\'e}thermin}, {Inami}, {Dickinson}, {Magnelli}, {Wang}, {Aussel}, {Daddi},
  {Juneau}, {Shu}, {Sargent}, {Buat}, {Faber}, {Ferguson}, {Giavalisco},
  {Koekemoer}, {Magdis}, {Morrison}, {Papovich}, {Santini}, \&
  {Scott}}]{Schreiber2015}
{Schreiber}, C., {Pannella}, M., {Elbaz}, D., {et~al.} 2015, \aap, 575, A74

\bibitem[{{Scoville} {et~al.}(2014){Scoville}, {Aussel}, {Sheth}, {Scott},
  {Sanders}, {Ivison}, {Pope}, {Capak}, {Vanden Bout}, {Manohar}, {Kartaltepe},
  {Robertson}, \& {Lilly}}]{Scoville2014}
{Scoville}, N., {Aussel}, H., {Sheth}, K., {et~al.} 2014, \apj, 783, 84

\bibitem[{{Scoville} {et~al.}(2016){Scoville}, {Sheth}, {Aussel}, {Vanden
  Bout}, {Capak}, {Bongiorno}, {Casey}, {Murchikova}, {Koda},
  {{\'A}lvarez-M{\'a}rquez}, {Lee}, {Laigle}, {McCracken}, {Ilbert}, {Pope},
  {Sanders}, {Chu}, {Toft}, {Ivison}, \& {Manohar}}]{Scoville2016}
{Scoville}, N., {Sheth}, K., {Aussel}, H., {et~al.} 2016, \apj, 820, 83

\bibitem[{{Shapley} {et~al.}(2003){Shapley}, {Steidel}, {Pettini}, \&
  {Adelberger}}]{Shapley2003}
{Shapley}, A.~E., {Steidel}, C.~C., {Pettini}, M., \& {Adelberger}, K.~L. 2003,
  \apj, 588, 65

\bibitem[{{Solomon} \& {Vanden Bout}(2005)}]{Solomon2005}
{Solomon}, P.~M. \& {Vanden Bout}, P.~A. 2005, \araa, 43, 677

\bibitem[{{Steidel} {et~al.}(2003){Steidel}, {Adelberger}, {Shapley},
  {Pettini}, {Dickinson}, \& {Giavalisco}}]{Steidel2003}
{Steidel}, C.~C., {Adelberger}, K.~L., {Shapley}, A.~E., {et~al.} 2003, \apj,
  592, 728

\bibitem[{{Tacconi} {et~al.}(2020){Tacconi}, {Genzel}, \&
  {Sternberg}}]{Tacconi2020}
{Tacconi}, L.~J., {Genzel}, R., \& {Sternberg}, A. 2020, \araa, 58, 157

\bibitem[{{Tacconi} {et~al.}(2013){Tacconi}, {Neri}, {Genzel}, {Combes},
  {Bolatto}, {Cooper}, {Wuyts}, {Bournaud}, {Burkert}, {Comerford}, {Cox},
  {Davis}, {F{\"o}rster Schreiber}, {Garc{\'\i}a-Burillo}, {Gracia-Carpio},
  {Lutz}, {Naab}, {Newman}, {Omont}, {Saintonge}, {Shapiro Griffin}, {Shapley},
  {Sternberg}, \& {Weiner}}]{Tacconi2013}
{Tacconi}, L.~J., {Neri}, R., {Genzel}, R., {et~al.} 2013, \apj, 768, 74

\bibitem[{{Valentino} {et~al.}(2021){Valentino}, {Daddi}, {Puglisi}, {Magdis},
  {Kokorev}, {Liu}, {Madden}, {G{\'o}mez-Guijarro}, {Lee}, {Cortzen},
  {Circosta}, {Delvecchio}, {Mullaney}, {Gao}, {Gobat}, {Aravena}, {Jin},
  {Fujimoto}, {Silverman}, \& {Dannerbauer}}]{Valentino2021}
{Valentino}, F., {Daddi}, E., {Puglisi}, A., {et~al.} 2021, \aap, 654, A165

\bibitem[{{Valentino} {et~al.}(2018){Valentino}, {Magdis}, {Daddi}, {Liu},
  {Aravena}, {Bournaud}, {Cibinel}, {Cormier}, {Dickinson}, {Gao}, {Jin},
  {Juneau}, {Kartaltepe}, {Lee}, {Madden}, {Puglisi}, {Sanders}, \&
  {Silverman}}]{Valentino2018}
{Valentino}, F., {Magdis}, G.~E., {Daddi}, E., {et~al.} 2018, \apj, 869, 27

\bibitem[{{Valentino} {et~al.}(2020){Valentino}, {Magdis}, {Daddi}, {Liu},
  {Aravena}, {Bournaud}, {Cortzen}, {Gao}, {Jin}, {Juneau}, {Kartaltepe},
  {Kokorev}, {Lee}, {Madden}, {Narayanan}, {Popping}, \&
  {Puglisi}}]{Valentino2020}
{Valentino}, F., {Magdis}, G.~E., {Daddi}, E., {et~al.} 2020, \apj, 890, 24

\bibitem[{{Vignati} {et~al.}(1999){Vignati}, {Molendi}, {Matt}, {Guainazzi},
  {Antonelli}, {Bassani}, {Brandt}, {Fabian}, {Iwasawa}, {Maiolino},
  {Malaguti}, {Marconi}, \& {Perola}}]{Vignati1999}
{Vignati}, P., {Molendi}, S., {Matt}, G., {et~al.} 1999, \aap, 349, L57

\bibitem[{{Wolfire} {et~al.}(2022){Wolfire}, {Vallini}, \&
  {Chevance}}]{Wolfire2022}
{Wolfire}, M.~G., {Vallini}, L., \& {Chevance}, M. 2022, \araa, 60, 247

\bibitem[{{Yang} {et~al.}(2017){Yang}, {Omont}, {Beelen}, {Gao}, {van der
  Werf}, {Gavazzi}, {Zhang}, {Ivison}, {Lehnert}, {Liu}, {Oteo},
  {Gonz{\'a}lez-Alfonso}, {Dannerbauer}, {Cox}, {Krips}, {Neri}, {Riechers},
  {Baker}, {Micha{\l}owski}, {Cooray}, \& {Smail}}]{Yang2017}
{Yang}, C., {Omont}, A., {Beelen}, A., {et~al.} 2017, \aap, 608, A144

\end{thebibliography}

\begin{appendix}

\section{Data of D49}
In Tables \ref{tab.data} and \ref{tab.mass} we list the photometric data we used in SED fitting and the \mgas\ estimates described in Section \ref{sec:gas_estimates}.

\begin{table}[htbp]
      \caption{Optical to millimetre photometry of D49.}
         \label{tab.data}
     $$ 
         \begin{array}{lll}
            \hline
            \hline
            \noalign{\smallskip}
            {\rm Band} & {\rm Flux\,[mJy]} \\
            \noalign{\smallskip}
            \hline
            \noalign{\smallskip}
            u{\rm \,Band}^{(a)} & 1.51 \pm 1.01 & \times 10^{-4}  \\
            g{\rm \,Band }^{(a)} & 1.24 \pm 0.21 & \times 10^{-3}  \\
            r{\rm \,Band}^{(a)} & 1.45 \pm 0.13 & \times 10^{-3}  \\
            J{\rm \,Band}^{(a)} & 2.15 \pm 0.36 & \times 10^{-3}  \\
            K{\rm \,Band}^{(a)} & 5.92 \pm 0.71 & \times 10^{-3}  \\
            \textit{Spitzer}\,{\rm IRAC}\,3.6\,{\rm \mu m}^{(a)} & 1.41 \pm 0.04 & \times 10^{-2}   \\
            \textit{Spitzer}\,{\rm IRAC}\,4.5\,{\rm \mu m}^{(a)} & 1.87 \pm 0.09 & \times 10^{-2}   \\
            \textit{Spitzer}\,{\rm IRAC}\,5.8\,{\rm \mu m}^{(a)} & 2.23 \pm 0.14 & \times 10^{-2}   \\
            \textit{Spitzer}\,{\rm IRAC}\,8.0\,{\rm \mu m}^{(a)} & 2.53 \pm 0.14 & \times 10^{-2}   \\
            \textit{Spitzer}\,{\rm MIPS}\,24\,{\rm \mu m}^{(b)} & 0.09 \pm 0.03   \\
            \textit{Herschel}\,{\rm PACS}\,100\,{\rm \mu m}^{(b)} & 4.27 \pm 1.22   \\
            \textit{Herschel}\,{\rm PACS}\,160\,{\rm \mu m}^{(b)} & 12.21 \pm 3.56   \\
            \textit{Herschel}\,{\rm SPIRE}\,250\,{\rm \mu m}^{(b)}& 15.08 \pm 1.80   \\
            \textit{Herschel}\,{\rm SPIRE}\,350\,{\rm \mu m}^{(b)} & 18.19 \pm 1.84     \\
            \textit{Herschel}\,{\rm SPIRE}\,500\,{\rm \mu m}^{(b)} & 11.83 \pm 2.37    \\
            {\rm IRAM}\,1.2\,{\rm mm}^{(b)} & 1.79 \pm 0.37   \\
            {\rm NOEMA}\,1.3\,{\rm mm}^{(c)} & 1.57 \pm 0.04 \\
            {\rm NOEMA}\,2.0\,{\rm mm}^{(c)} & 0.30 \pm 0.03 \\
            {\rm NOEMA}\,2.3\,{\rm mm}^{(c)} & 0.21 \pm 0.03 \\
            \noalign{\smallskip}
            \hline
         \end{array}
     $$ 
     \tablefoottext{a}{\cite{Rigopoulou2006}}
     \tablefoottext{b}{\cite{Magdis2017}}
     \tablefoottext{c}{This work}
\end{table}

\begin{table}[htbp]
      \caption{Hydrogen mass estimations by different methods.}
         \label{tab.mass}
     $$ 
         \begin{array}{cc}
            \hline
            \hline
            \noalign{\smallskip}
            {\rm Method} & {\rm Estimated}\,\mathrm{log}(M_{\rm gas}/M_\odot)\\
            \noalign{\smallskip}
            \hline
            \noalign{\smallskip}
            {\rm CO}\,(\langle r_{31}\rangle = 0.5\pm0.15, \alpha_{\rm CO} = 3.5) & 11.48 \pm 0.23^{(a)} \\
            \delta_{\rm GD} Z_\odot & 11.12 \pm 0.25^{(a)} \\
            \delta_{\rm GD} {\rm "broken"FMR} & 11.34 \pm 0.25^{(a)} \\
            {\rm R-J} & 11.29 \pm 0.31^{(a)} \\
            {\tt Stardust} {\rm \,\, fitting} & 11.19 \pm 0.17^{(b)} \\
            {\rm [CI] pair\,(non-LTE)} & < 11.22^{(b)} \\
            {\rm [CI](1-0)} \,(\alpha_{\rm CI} = 17.0) & < 11.16^{(b)} \\
            \noalign{\smallskip}
            \hline
         \end{array}
     $$ 
     \tablefoottext{a}{\cite{Magdis2017}}
     \tablefoottext{b}{This work}
   \end{table}

\end{appendix}

\end{document}